\documentclass[12pt,a4paper]{article}

\usepackage{natbib} % <-- new
\usepackage[latin1]{inputenc}
\usepackage{authblk}    % per blocco autori e affiliazioni
\usepackage{makeidx}         % allows index generation
\usepackage{graphicx}        % standard LaTeX graphics tool
\usepackage{amsmath}
\usepackage{amsthm}
\usepackage{amsfonts}
\usepackage{appendix}
\usepackage{url}
\usepackage{subcaption} %double figures
\usepackage{algorithm}
\usepackage{algpseudocode}
\usepackage{adjustbox} %scale figure
\usepackage{bbm}%indicator function

\usepackage{array}
\newcolumntype{C}[1]{>{\centering\arraybackslash}m{#1}}

\usepackage{tikz-cd}    %hiddenmarkovchain
\usetikzlibrary{backgrounds}

\usepackage[colorlinks=true, allcolors=blue]{hyperref}

\usepackage[tikz]{bclogo}
\usetikzlibrary{calc}
\usepackage[font={small}]{caption}

\newcommand{\mailadd}{\href{mailto:me@somewhere.com}{\vspace*{0.15cm}}} 

\usepackage{threeparttable}
 

\title{\textbf{Latent event history models for quasi-reaction systems}
}

\author[a]{Matteo Framba}
\author[a]{Veronica Vinciotti}
\author[b]{Ernst Wit}
\affil[a]{\small University of Trento; \mailadd{\texttt{matteo.framba@unitn.it}}, \mailadd{\texttt{veronica.vinciotti@unitn.it}}}
\affil[b]{\small Universit\`a della Svizzera italiana; \mailadd{\texttt{ernst.jan.camiel.wit@usi.ch }
 }}
\date{}
\begin{document}
\maketitle
\begin{abstract}
Various processes can be modelled as quasi-reaction systems of stochastic differential
equations, such as cell differentiation and disease spreading. Since the underlying data of
particle interactions, such as reactions between proteins or contacts between people, are
typically unobserved, statistical inference of the parameters driving these systems is developed from concentration data measuring each unit in the system over time.
While observing the continuous time process at a time scale as fine as possible should
in theory help with parameter estimation, the existing Local Linear Approximation (LLA)
methods fail in this case, due to numerical instability caused by small
changes of the system at successive time points. On the other hand, one
may be able to reconstruct the underlying unobserved interactions from the observed count
data. Motivated by this, we first formalise the latent event history model underlying the
observed count process. We then propose a computationally efficient Expectation-Maximation algorithm for parameter estimation, with an extended Kalman filtering procedure for the prediction of the latent states. 
A simulation study shows the performance of the proposed method and highlights the settings where it is particularly advantageous compared to the existing LLA approaches. Finally, we present an illustration of the methodology on the spreading of the COVID-19 pandemic in Italy.
\end{abstract}



\textbf{Keywords:} SDEs, Euler-Maruyama, Kalman Filter, EM algorithm



\section{Introduction}
\label{sec:1}
An increasing number of natural phenomena can be described by quasi-reaction systems of stochastic differential equations, as these are able to capture the inherent stochasticity of many processes.  Examples include the stem cell differentiation process \citep{pellin19, pellin22}, the dynamics of a biological system \citep{wilkinson18} or of an infectious disease spreading \citep{britton19}, and the diverse applications of diffusion processes \citep{craigmile23}.  The dynamics of these systems depend critically on parameters which are often unknown. Estimating these parameters is therefore important for characterizing and predicting the evolution of a dynamic system. 

In this framework, the likelihood of the intermittently observed process has rarely an explicit form \citep{wilkinson18}.
To overcome this problem, Local Linear Approximation (LLA) methods provide an explicit approximation of the likelihood function under some assumptions \citep{Shoji}. Nevertheless, both in the case when observations are too spaced out in time and when the inter-observations times are too close, estimates based on the LLA are biased. \cite{Komorowski} present an extensive study on the effects of correlation between molecule concentrations  on statistical inference, in the specific case of stochastic chemical kinetics models. 
Various approaches for reducing the variance of parameter estimators  in a generic multi-response, non-linear model are available and could be used also in the case of dynamic systems. In the context of D-optimal designs, the most commonly used criterion assumes knowledge of the variance-covariance matrix \citep{fedorov2013theory}.  Although alternatives exist that use only an estimate of this matrix \citep{cooray1987sequential}, recent studies have observed that minimising the determinant of the information matrix is computationally efficient but not very robust  \citep{hatzis1992optimal}.
An alternative  approach is the use of Tikhonov  regularisation techniques \citep{engl96}. However, if the measurements are taken very close together in time, the concentrations can be constant, leading to zero standard deviations and making also regularisation infeasible. %: in fact, a pre-process of standardisation is necessary to prevent dominance of variables with larger magnitudes over the others as the scales of the responses are  often of different orders. 

In this paper, we propose an approach to overcome these limitations.  Intuitively, when the concentrations of the particles in the system are observed very close in time, one may be able to reconstruct which events of the stochastic process have taken place in order to result in a change of the system from the current to the next time point. Thus, the core element of our proposed approach involves integrating event history analysis into the framework of quasi-reaction systems. Originally conceived for sociological studies, event history models have been used on a range of applications, from engineering to medicine, economics, political science and psychology  \citep{Box2}.  %Rather than focusing on the evolution of the states, these models focus on the events themselves whose occurrence changes the states. 
As the rates governing the evolutions of the state of the system and of the underlying event counting process are clearly linked, and they depend on the previous state of the system, the first contribution of the paper will be to formalise a joint statistical model that couples the two processes. 

The second contribution of the paper is to develop an inferential procedure for the proposed model. As the occurrence of the events is not observed, we derive an Expectation-Maximation (EM) algorithm for parameter estimation. In particular, at the E-step,  Kalman filter \citep{kalman60} is used for the prediction of the latent events from the dynamics of the system observed on the entire time interval. The most popular version of Kalman filtering is for the case of Gaussian linear systems. However, an extension is required for our purposes. Firstly, since the latent event counts are not Gaussian, we propose to approximate the Poisson distribution with a continuous Gamma distribution. The latter is then transformed to a Gaussian distribution via a marginal transformation. Secondly, since the resulting system is non-linear, we propose an extended Kalman filtering procedure for estimating  the latent state of the event count process \citep{Anderson}. This allows the evaluation of the Q-function, which is then maximised at the M-step of the EM algorithm. In this way, our approach relates to other implementations of EM algorithms with an embedded Kalman filter, such as  \citep{shumway1982approach,ghahramani1996parameter} for dynamic linear systems, and more recent extensions for non-linear systems, such as the EM extended Kalman filter of \cite{bar2001estimation}, the EM unscented Kalman filter of \cite{wan2000unscented} and the  EM particle filter of \cite{zia2008algorithm}.

The  rest of the paper is organized as follows. In Section \ref{sec:2}, we formalise the proposed latent event history model for quasi-reaction systems. In Section \ref{sec:3}, we describe the EM algorithm for parameter estimation.  In Section \ref{sec:5}, a simulation study shows the performance of the proposed method and highlights the settings where it is particularly advantageous compared to the existing LLA approaches.  In Section \ref{sec:6}, we describe an illustration of the method on the modelling of the COVID-19 transmission dynamics in Italy. Finally, in Section \ref{sec:conclusion}, we draw some conclusions and point to directions for future work.
 
%%%%%%%%%%%%%%%%%%%%%%%%%%%%%%%%%%%%
\section{Modelling quasi-reaction systems} \label{sec:2}
Consider a closed system in which $p$ substrates interact, each denoted as $Q_l$ with $l=1,\dots,p$. These substrates could represent the compartments of an infectious disease model, the cell types in a cell differentiation model, or the different molecules in a biochemical reaction system. The $j$-th chemical reaction can generally be described as
\begin{equation}
	\label{initial}
	k_{1j}Q_1+... +k_{pj}Q_p \xrightarrow{\beta_j}  s_{1j}Q_1+... +s_{pj}Q_p \quad\quad j\in1,\dots,r,
\end{equation}
where $r$ indicates the number of reactions describing the dynamic system. We refer to $\mathcal{R}$ as the set of possible reactions.  The \textit{stoichiometric coefficients} $k_{lj}$ and $s_{lj}$ are fixed integer values that describe the amount of substrate $l$, as reactant and product, respectively, that is needed for reaction $j$ to occur, while  $\theta_j=\exp(\beta_j) \in \mathbb{R}^{+}$ is the rate at which reaction $j$ occurs.  

The log-reaction rates $\boldsymbol\beta = (\beta_1,\ldots,\beta_r)^{\top}$ characterize the evolution of the dynamic system. These are the parameters that need to be estimated, given realizations of the state of the system over time.
Let then $Y_l(t)$ denote the concentration of the $l$-th particle at time $t$, with $t\in[0,T]$. Let $\textbf{Y}(t)=(Y_{1}(t),\dots,Y_{p}(t))^T\in \mathbb{N}_0^p$ denote the state of the system at time $t$.
%The biological system is sometimes identified as a undirected graph, where the nodes represent the reactants and the edges the reactions that may occur. With the appropriate assumptions, networks of coupled chemical reactions are specific enough in their description of the system to allow both simulations of the dynamics and inferential studies of possible weights on the edges \citep{wilkinson18}.
Even if reaction equations like (\ref{initial}) are often used to represent kinetic models, as these facilitate a qualitative understanding of the dynamics, from a mathematical point of view chemical reactions are modelled primarily as systems of stochastic differential equations \citep{wilkinson18}. This methodology enables a quantitative interpretation of the dynamics, as it allows to study the temporal variation of the concentrations $\textbf{Y}$  from the dynamics at the unit level.

Two viewpoints can be taken in the characterization of the stochastic process that induces changes in the state $\textbf{Y}$ over time. According to the underlying dynamic system, particles come to interact at a particular point in time, reactions are fired instantaneously as a result of this and the system moves to the next state. In the first viewpoint, which we present in Section \ref{sec:2.1}, one can study the stochastic process by which reactions occur and, thus, the state of the system $\textbf{Y}$ moves to a new configuration. A second viewpoint, which we describe in Section  \ref{sec:2.2}, is to describe directly the change of the system based on the amount of particles available at a certain point in time and the hazard rate of each reaction at that point in time. As the occurrence of reactions is not observed, the second viewpoint is the most direct approach for modelling dynamic systems. Indeed, this is the approach considered in the literature and the one that results in the traditional LLA approaches for parameter estimation. As it was argued in the introduction, particularly when the system is observed at small time intervals, one may be able to reconstruct the underlying process of reactions, leading to a more accurate characterization of the dynamic system, particularly in the case of high temporal correlation between the states. Motivated by this, Section \ref{sec:2.3} shows how the two viewpoints can be unified into a joint statistical model. 

%%%%%%%%%%%%
\subsection{Event process}  \label{sec:2.1}
Let $e_{j}:=(t_j, r_{j})$ be the \textit{event} that reaction $j\in \mathcal{R}$ occurs at time $t_j$. Associated with this marked point process and with each reaction $j$, we define the multivariate counting process
\begin{equation*}
N_j(t)=\#\{\text{Reactions of type $j$ occurring in time interval } [0,t], j\in \mathcal{R}\}.
\end{equation*}
We assume that $N_j(t)$ follows a non-homogeneous Poisson process
\[N_{j}(t) \sim \mbox{Poisson}(\Lambda_{j}(t)),\]
with cumulative rate
\[\Lambda_{j}(t)=E[N_{j}(t) \:|\: {\cal F}_{t^-}]=\int_0^t\lambda_{j}(\boldsymbol{Y}(u);\boldsymbol{\beta})du,\]
where ${\cal F}_{t^-}$ is the history of the process up to, but excluding, time $t$.

The hazard rate $\lambda_{j}(\boldsymbol{Y}(t);\boldsymbol{\beta})$ depends on the state of the system at time $t$ as well as on the amount of particles of each type that are needed for each reaction to occur, i.e., the stoichiometric coefficients $k_{lj}$ in (\ref{initial}). In particular, assuming that the times at which reactions occur are exponentially distributed and independent of each other, we have that
\begin{equation}
	\label{hazard}
	\lambda_j(\boldsymbol{Y}(t);\boldsymbol{\beta}) = \exp({{\beta}_j}) \prod_{l=1}^p  {{Y}_{l}(t)\choose k_{lj}},
\end{equation}
where ${Y_{l}(t) \choose k_{lj}}=0,$  for all ${{Y}_{l}(t) <k_{lj}}$. 

\subsection{Particle concentration process}   \label{sec:2.2}
The state of the system $\boldsymbol{Y}(t)$ is itself also a continuous time discrete Markov process. In the particular setting of a quasi-reaction system, it is possible to establish the temporal evolution of the probability distribution $P(\textbf{Y};t)$, i.e., the probability that $\textbf{Y}$ is the state of the system at time $t$. This will again depend on the state of the system just before time $t$. In particular, the distribution can be shown to satisfy the chemical master equations \citep{wilkinson18}
\begin{equation}
	\label{master}
	\frac{d P(\boldsymbol{Y} ; t)}{d t}=\sum_{j\in\mathcal{R}}\left[\lambda_j\left(\boldsymbol{Y}(t)-V_{\cdot,j};\boldsymbol{\beta}\right) P\left(\boldsymbol{Y}-V_{\cdot,j} ; t\right)-\lambda_j(\boldsymbol{Y}(t);\boldsymbol{\beta}) P(\boldsymbol{Y} ; t)\right],
\end{equation}
where $V$ denotes the net effect matrix,  with $(l,j)$ entry given by $v_{lj}=s_{lj}-k_{lj}$.

A solution of (\ref{master}) gives the full transition probability kernel for the system dynamics. The master equations, however, can be solved analytically only in a small number of cases, due to the vast spectrum of conceivable state configurations \citep{mcquarrie1967stochastic}. On the other hand, from the master equations, one can derive the conditional expectation and variance of the rate of changes of the system. These are given, respectively, by 
\begin{align}
\label{EMapprox}
&\frac{\mathbb{E}[\textbf{Y}(t+dt)-\textbf{Y}(t) \:|\: \textbf{Y}(t)]}{dt}= V\boldsymbol{\lambda}(\boldsymbol{Y}(t);\boldsymbol{\beta}) \\ \nonumber
&
\frac{\mathbb{V}[\textbf{Y}(t+dt)-\textbf{Y}(t) \:|\: \textbf{Y}(t)]}{dt}= V\text{diag}(\boldsymbol{\lambda}(\boldsymbol{Y}(t);\boldsymbol{\beta}))V^T.
\end{align}
These two moments can be used as the basis of an approximate solution to the master equations.

%%%%%%%%%%%%%%%%%%%%%%%%%%%%%%%%%%%%%%%%%
\subsection{Latent event history model \label{sec:2.3}}

We now merge the two characterizations described above into one joint model based on realizations of the process at discrete time points. Let then $\textbf{Y}_{i}=\textbf{Y}(t_i)$, $i= 0,\ldots,N$, be the state of the process at $N+1$, not necessarily equispaced, time points. Under the non-homogeneous Poisson process described in Section \ref{sec:2.1}, and assuming that the hazard rates remain constant within the $N$ time intervals, we have that
\begin{equation}
N_j(t_i)-N_j(t_{i-1}) \:|\: \mathcal{F}_{t_{i-1}}\sim \mbox{Poisson}(\mu_{ij}(\textbf{Y}_{i-1};\boldsymbol{{\beta}})), \quad j=1,\ldots, r,
\label{poi}
\end{equation}
where 
\begin{equation}
	\label{mu2}
	\mu_{ij}(\textbf{Y}_{i-1};\boldsymbol{{\beta}})= (t_{i}-t_{i-1})\lambda_j(\textbf{Y}_{i-1};\boldsymbol{{\beta}}),
\end{equation}
with $\lambda_j(\textbf{Y}_{i-1};\boldsymbol{{\beta}})$ defined as in Equation (\ref{hazard}).
For the rest of the manuscript, we simplify the notation and denote with $\boldsymbol{\mu}_i$ the vector $\boldsymbol{\mu}_i(\textbf{Y}_{i-1};\boldsymbol{{\beta}})$ of rates for all $r$ reactions at time $t_i$. 

In view of applying Gaussian state space models, we consider an approximation of the Poisson distribution with a continuous distribution, which we then transform to a Gaussian distributed random variable via a marginal transformation. In particular,  we associate the random variable $X_{ij}=F_{ij}^{-1}(\Phi_{ij}(Z_{ij}))$ to the increments $N_j(t_i)-N_j(t_{i-1})$, related to reaction $j$ and the time interval $(t_{i-1}, t_i]$. We define $X_{ij}$ by setting $F_{ij}$ as the CDF of a Gamma distribution with scale parameter $1$ and shape parameter $\mu_{ij}(\textbf{Y}_{i-1};\boldsymbol{{\beta}})$ from Equation (\ref{mu2}). This way both the mean and variance of $X_{ij}$ match that of the Poisson increments. Furthermore, it has a similar skewness.  $\Phi_{ij}$ is the CDF of a Gaussian distribution with mean and variance both equal to $\mu_{ij}(\textbf{Y}_{i-1};\boldsymbol{{\beta}})$. So $Z_{ij}$ is the Gaussian random variable that is uniquely associated to the Gamma distributed random variable $X_{ij}$, and with the same conditional mean and variance as the original $N_j(t_i)-N_j(t_{i-1})$ variable. In the remaining of the paper, we denote with $\textbf{Z}_i$ the $r$-dimensional vector of Gaussian random variables associated to the event counts in the interval $(t_{i-1}, t_i]$, $\textbf{X}_i$ the corresponding Gamma random variables and  $G$ the function that transforms $\textbf{Z}_i$ into $\textbf{X}_i$, namely 
\[\textbf{X}_i= G(\textbf{Z}_i)=\big(F_{i1}^{-1}(\Phi_{i1}(Z_{i1})),\ldots,F_{ir}^{-1}(\Phi_{ir}(Z_{ir}))\big).\]

The observed process $\textbf{Y}_i$ is also discrete and we apply also here an approximation to a Gaussian process. Firstly, using the assumption that the process is homogeneous within the time intervals, Equations (\ref{EMapprox}) result in 
\begin{align*}
&\mathbb{E}[\textbf{Y}_i-\textbf{Y}_{i-1} \:|\: \textbf{Y}_{i-1}]= V\boldsymbol{\mu}_i \\ 
&\mathbb{V}[\textbf{Y}_i-\textbf{Y}_{i-1} \:|\: \textbf{Y}_{i-1}]= V\text{diag}(\boldsymbol{\mu}_i)V^T.
\end{align*}
Secondly, we apply a Euler-Maruyama approximation of the discrete state process  into a continuous state diffusion process, characterized by the same conditional means and variances. Thus, we assume
\begin{equation}
	\label{gaussian}
\textbf{Y}_i \: | \: \textbf{Y}_{i-1}\sim \mathcal{N}\bigg( \textbf{Y}_{i-1}+ V\boldsymbol{\mu}_i, V{\text{diag}(\boldsymbol{\mu}}_i)V^T  \bigg).
\end{equation}

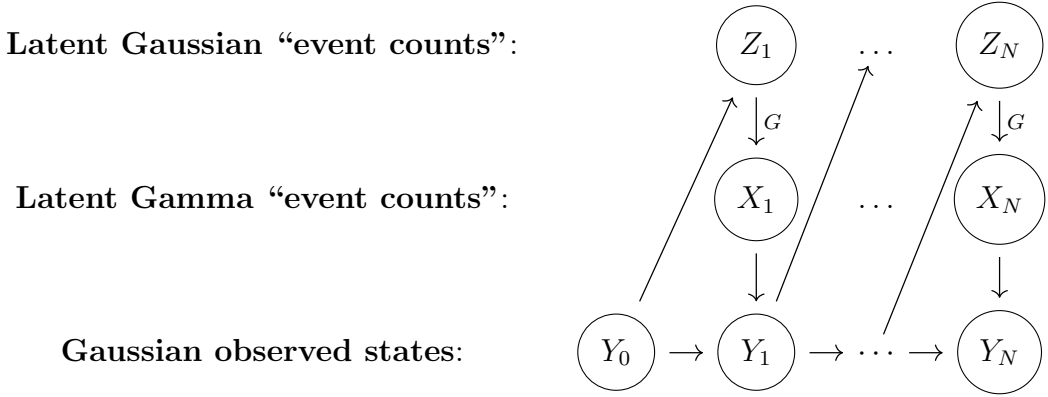
\begin{figure}[t!]
	\centering
	\begin{tikzcd}[nodes in empty cells, row sep=0.3em,column sep=1em]
		\text{\textbf{Latent Gaussian ``event counts''}:}
		& 
		&   \tikz{\node[circle,draw,minimum size=1cm] (X1) at (0,0) {$Z_1$};}  \ar[dd, "G"]
		&   \dots 
		&   \tikz{\node[circle,draw,minimum size=1cm] (XN1) at (0,0) {$Z_{N}$};} \ar[dd, "G"]     \\
		&     \\
		\text{\textbf{Latent Gamma ``event counts''}:}
		& 
		&   \tikz{\node[circle,draw,minimum size=1cm] (X1) at (0,0) {$X_1$};}  \ar[dd]
		&   \dots 
		&   \tikz{\node[circle,draw,minimum size=1cm] (XN1) at (0,0) {$X_{N}$};}  \ar[dd]       \\
		&     \\
		\text{\textbf{Gaussian observed states}:}
		&   \tikz{\node[circle,draw,minimum size=1cm] (Y0) at (0,0) {$Y_0$};} \ar[ruuuu] \ar[r]
		&   \tikz{\node[circle,draw,minimum size=1cm] (Y1) at (0,0) {$Y_1$};}\ar[ruuuu] \ar[r]
		&   \dotsm \ar[ruuuu]\ar[r]
		&   \tikz{\node[circle,draw,minimum size=1cm] (YN1) at (0,0) {$Y_{N}$};}      
	\end{tikzcd}
	\caption{\textbf{Latent event history model DAG}. The state of the system at time $t_i$, $\textbf{Y}_i$, depends on the previous state, $\textbf{Y}_{i-1}$, and on the number of reactions $\textbf{X}_i$ that occur in the interval $(t_{i-1},t_i]$.  The latent $\textbf{X}_i$ is approximated with a Gamma distribution and connected deterministically with a Gaussian random vector $\textbf{Z}_i$, via a marginal transformation $G$.  Notice how $\textbf{Z}_i$ is independent of future states, $\textbf{Y}_{(i+1):N}$, conditional on current and past states, $\textbf{Y}_{0:i}$.}
	\label{directed_approx}
\end{figure}

The models (\ref{poi}) and (\ref{gaussian}), associated to the two viewpoints described in Section \ref{sec:2.1} and \ref{sec:2.2}, respectively, are clearly connected by shared parameters. Moreover, it is clear how knowledge of the event counts $\textbf{X}_i$ in the interval $(t_{i-1},t_i]$, and equivalently of $\textbf{Z}_i$, would allow for perfect prediction of the state of the system at time $t_i$, since $\textbf{Y}_i-\textbf{Y}_{i-1}= V	\textbf{X}_i$. In the following, we will formulate the model more generally, so as to account also for possible measurement error in the observations $\textbf{Y}_i$, which may be relevant in some applied settings.  Putting all things together, we propose in this paper the following latent event history model
\begin{equation}
	\left\{
	\begin{aligned}
		\textbf{Z}_{i} &= \boldsymbol{\mu}_i+\boldsymbol{\varepsilon}_{i},  &\boldsymbol{\varepsilon}_{i}\sim&\mathcal{N}\big(0,\text{diag}(\boldsymbol{\mu}_i)\big),\\
			\textbf{Y}_{i} &= \textbf{Y}_{i-1}+V	G(\textbf{Z}_i)  +\boldsymbol{\psi}_i,  &\boldsymbol{\psi} _i \sim& \mathcal{N}(0,\Sigma), i=1,\dots,N
	\end{aligned}
	\right.
\label{statespace2}
\end{equation}
where  $\boldsymbol{\psi}_i$ is a Gaussian noise vector with mean zero and variance-covariance $\Sigma=\sigma^2\cdot \mathbb{I}_p$. The case of no measurement error in $\textbf{Y}_i$, which we will consider in the simulations and real data illustration, will correspond to the special case of $\sigma^2=0$.  Figure \ref{directed_approx} summarizes the dependence structure associated to the proposed model.

\section{Inference} \label{sec:3}
In this section, we discuss statistical inference of the latent event history model (\ref{statespace2}). Denoting with $\textbf{Y}$ the $(N+1)\times p$ matrix of observations at the $N+1$ time points and $\textbf{Z}$ the $(N+1)\times r$ matrix of latent variables,  estimation of $\boldsymbol{{\beta}}$ requires the optimization of the marginal log-likelihood 
\begin{equation}
\ell_{\textbf{Y}}(\boldsymbol{{\beta}}) =\log \int_\textbf{Z} L_{\textbf{Z}, \textbf{Y}}(\boldsymbol{{\beta}})d\textbf{Z}.
\label{eq:marglik}
\end{equation}
As common in the presence of latent variables, we derive an Expectation-Maximisation (EM) algorithm for parameter estimation \citep{Dempster}. 
To this end, the complete log-likelihood, conditional on the initial state $\textbf{Y}_0$ and assuming some measurement error in $\textbf{Y}_i$ ($\sigma^2 \ne 0$), can be factorized into
\begin{equation}
	\label{full}
	\ell_{\textbf{Z}, \textbf{Y}}(\boldsymbol{\beta}) = \sum_{i=1}^{N} \big[\ell_{\textbf{Z}_i| \textbf{Y}_{i-1}}(\boldsymbol{\beta})+\ell_{ \textbf{Y}_i|\textbf{Z}_i,\textbf{Y}_{i-1}}\big],
\end{equation}
where
\begin{equation}
	\label{CP}
	\ell_{\textbf{Z}_i| \textbf{Y}_{i-1}}(\boldsymbol{\beta}) =-\frac{1}{2}\biggl\{r\log(2\pi)+\log(|\text{diag}(\boldsymbol{\mu}_i)|)+\big[\textbf{Z}_i-\boldsymbol{\mu}_i\big]^T(\text{diag}(\boldsymbol{\mu}_i))^{-1}\big[\textbf{Z}_i-\boldsymbol{\mu}_i\big]\biggl\}
	\notag
\end{equation}
and
\begin{align}
	\label{LLA}
	\ell_{ \textbf{Y}_i|\textbf{Z}_i,\textbf{Y}_{i-1}}= &-\frac{1}{2}\biggl\{p\log(2\pi)+\log(|\Sigma|) \notag \\
	&+\big[ \textbf{Y}_i-\textbf{Y}_{i-1}-VG(\textbf{Z}_i)\big]^T\Sigma^{-1}\big[ \textbf{Y}_i-\textbf{Y}_{i-1}-VG(\textbf{Z}_i)\big]\biggl\}.
	\notag
\end{align}

The EM algorithm will then consist in the following two steps, which are iterated until convergence:
\begin{itemize}
\item \textit{E-step}: Setting $\boldsymbol{{\beta}}$ to the current estimate  of the parameters,  $\boldsymbol{\beta}^*$,  compute the expected value of the complete log-likelihood (\ref{full}) with respect to the distribution of the latent variables given the observations:
\begin{equation}
	\label{Q}
	Q(\boldsymbol{{\beta}}|\boldsymbol{{\beta}}^*)
	=\mathbb{E}_{\textbf{Z}|\textbf{Y},\boldsymbol{\beta}^*}[\ell_{\textbf{Z}, \textbf{Y}}(\boldsymbol{{\beta}})].
\end{equation}
\item \textit{M-step}: 
  Find the optimal $\boldsymbol{\beta}$ by maximising the objective function (\ref{Q}) with respect to $\boldsymbol{{\beta}}$.
\end{itemize}
In the next two sections, we discuss in detail the computational aspects associated to the two steps, respectively.

\subsection{E-step: Extended Kalman filtering}
With the complete log-likelihood written as in (\ref{full}), the Q-function (\ref{Q}) with slight abuse of notation is given by
 \begin{equation}
Q(\boldsymbol{{\beta}}|\boldsymbol{{\beta}}^*)=\mathbb{E}[\ell_{\textbf{Z}| \textbf{Y} }(\boldsymbol{{\beta}})|\textbf{Y}_{0:N}]+\mathbb{E}[\ell_{ \textbf{Y}|\textbf{Z}\textbf{Y} }|\textbf{Y}_{0:N}], \label{Qfun}
 \end{equation}
where $\textbf{Y}_{0:N}$ denotes the data across all time points. Under model (\ref{statespace2}), the first term involves the following expectation
\begin{align*}
	\mathbb{E}[\ell_{\textbf{Z}| \textbf{Y} }(\boldsymbol{{\beta}})|\textbf{Y}_{0:N}]=&\mathbb{E}\bigg[-\frac{1}{2}\sum_{i=1}^{N}\biggl\{r\log(2\pi)+\log(|\text{diag}({\boldsymbol{\mu}}_i)|) \notag \\
	&+\big[\textbf{Z}_i-{\boldsymbol{\mu}}_i\big]^T\text{diag}({\boldsymbol{\mu}}_i)^{-1}\big[\textbf{Z}_i-{\boldsymbol{\mu}}_i\big]\biggl\}|\textbf{Y}_{0:N}\bigg]\notag\\
%& \propto 
%-\frac{1}{2}Nr\log(2\pi) \!-\!\frac{1}{2}\sum_{i=1}^{N}\bigg\{ \log(|\text{diag}({\boldsymbol{\mu}}_i)|)+
% \!-\!\frac{1}{2}\sum_{i=1}^{N}\bigg\{ \mathbb{E}\big[  \textbf{Z}_i^T\text{diag}({\boldsymbol{\mu}}_i)^{-1}\textbf{Z}_i
%-2\textbf{Z}_i^T\text{diag}({\boldsymbol{\mu}}_i)^{-1} \boldsymbol{\mu}_i 
%+\boldsymbol{\mu}_i^T\text{diag}({\boldsymbol{\mu}}_i)^{-1}\boldsymbol{\mu}_i |\textbf{Y}_{0:N}\big]\bigg\}\notag\\
 \propto & 
%-\frac{1}{2}Nr\log(2\pi)\!-\!\frac{1}{2}\sum_{i=1}^{N}\bigg\{ \log(|\text{diag}({\boldsymbol{\mu}}_i)|)
- \frac{1}{2}\sum_{i=1}^{N}\bigg\{\mathbb{E}\big[  \textbf{Z}_i^T|\textbf{Y}_{0:N}\big]\text{diag}({\boldsymbol{\mu}}_i)^{-1}\mathbb{E}\big[\textbf{Z}_i|\textbf{Y}_{0:N}\big] \notag \\
&+\text{Tr}\big[\text{diag}({\boldsymbol{\mu}}_i)^{-1}\mathbb{V}[\textbf{Z}_i|\textbf{Y}_{0:N}]\big] -2\mathbb{E}\big[\textbf{Z}_i^T|\textbf{Y}_{0:N}\big]\!\cdot\!\boldsymbol{1}+\boldsymbol{\mu}_i^T\!\cdot\!\boldsymbol{1}  \bigg\}, \notag
\end{align*}
while expectation of the second term results in
\begin{align*}
	\mathbb{E}[\ell_{ \textbf{Y}|\textbf{Z}\textbf{Y} }|\textbf{Y}_{0:N}]=&\mathbb{E}\bigg[\!-\frac{1}{2}\sum_{i=1}^{N} \biggl\{p\log(2\pi)+\log(|\Sigma|) \notag \\
	&+\big[\Delta \textbf{Y}_i-VG(\textbf{Z}_i)\big]^T\Sigma^{-1}\big[\Delta \textbf{Y}_i-VG(\textbf{Z}_i)\big]\biggl\}|\textbf{Y}_{0:N}\bigg]\notag\\
	%&\propto
	%-\frac{1}{2}N p \log (2 \pi)	-\frac{1}{2}N \log (\sigma^{2p})	-\frac{1}{2\sigma^{2}}\sum_{i=1}^{N}\bigg\{\mathbb{E}\bigg[\Delta \textbf{Y}_i^{T} \Delta \textbf{Y}_i \mid \textbf{Y}_{0:N}\bigg]
	%-\frac{1}{2\sigma^{2}}\sum_{i=1}^{N}\bigg\{-2 \mathbb{E}\bigg[\Delta \textbf{Y}_i^{T} V G(\textbf{Z}_i) \mid \textbf{Y}_{0:N}\bigg]+\mathbb{E}\bigg[\left(V G\big(\textbf{Z}_i\big)\right)^{T}\big(V G(\textbf{Z}_i)\big) \mid \textbf{Y}_{0:N}\bigg]\bigg\}\notag\\
	\propto &
	%=-\frac{1}{2}N p \log \left(2 \pi \sigma^2\right)	
	- \frac{1}{2\sigma^{2}}\sum_{i=1}^{N}\bigg\{\!-\!2\Delta \textbf{Y}_i^{T} V  \mathbb{E}\big[G(\textbf{Z}_i)|\textbf{Y}_{0:N}\big]\notag \\
	&+\mathbb{E}\big[G(\textbf{Z}_i)^T|\textbf{Y}_{0:N}\big]V^TV\mathbb{E}\big[G(\textbf{Z}_i) \mid \textbf{Y}_{0:N}\big] \notag \\
	&+\text{Tr}(V^{T} V \mathbb{V}\big[G(\textbf{Z}_i)|\textbf{Y}_{0:N}\big])\bigg\},\notag
\end{align*}
with $\Delta \textbf{Y}_i= \textbf{Y}_i-\textbf{Y}_{i-1}$ and where we retain for the moment only the terms dependent on the latent variables. 

In particular, we can see how the calculation of the Q-function requires the evaluation of the following first and second moments: $\mathbb{E}[\textbf{Z}_i|\textbf{Y}_{0:N}] $, $\mathbb{V}[\textbf{Z}_i|\textbf{Y}_{0:N}]$, $\mathbb{E}[G(\textbf{Z}_i)|\textbf{Y}_{0:N}] $, and $\mathbb{V}[G(\textbf{Z}_i)|\textbf{Y}_{0:N}]$. 
To this end, we use a Kalman filter approach. Firstly, notice how the dependences implied by model (\ref{statespace2}) are such that
\begin{align*}
\mathbb{E}[\textbf{Z}_i|\textbf{Y}_{0:N}] &= \mathbb{E}[\textbf{Z}_i|\textbf{Y}_{0:i}]\\
\mathbb{V}[\textbf{Z}_i|\textbf{Y}_{0:N}]&=\mathbb{V}[\textbf{Z}_i|\textbf{Y}_{0:i}]\\
\mathbb{E}[ G(\textbf{Z}_i)|\textbf{Y}_{0:N}]	&= \mathbb{E}[ G(\textbf{Z}_i)|\textbf{Y}_{0:i}]\\
\mathbb{V}[ G(\textbf{Z}_i)|\textbf{Y}_{0:N}] &=	\mathbb{V}[G(\textbf{Z}_i)|\textbf{Y}_{0:i}],
\end{align*}
since $\textbf{Z}_i$ is independent of future states, $\textbf{Y}_{(i+1):N}$, conditional on current and past states, $\textbf{Y}_{0:i}$ (Figure \ref{directed_approx}). In the following, we denote the first two quantities with $\hat{\textbf{z}}_{i|i}$ and $V_{i|i}$, respectively.  This means that the smoothing step of a traditional Kalman filtering procedure is not needed, and only the prediction and update steps are. Secondly, the non-linearity in $\textbf{Z}_i$ induced by the marginal transformation $G$ means that a standard Kalman filter approach is not applicable. Thus, in order to calculate the first and second moments of $G(\textbf{Z}_i)$, we consider an  extended Kalman filter, where we  approximate the function $G$ with a second order Taylor expansion.

According to the derivations in Appendix \ref{app:estep}, we find that the first two expectations are given by
\begin{align*}
\hat{\textbf{z}}_{i \mid i} &=\mathbb{E}\left[\textbf{Z}_{i} \mid \textbf{Y}_{0: i}\right]=\hat{\textbf{z}}_{i\mid i-1}+K_{i}\bigg[\textbf{Y}_{i}-\textbf{Y}_{i-1}-V\big(\textbf{g}_{i|i-1}+\frac{1}{2}\text{vect}(V_{i|i-1}H_{i|i-1})\big)\bigg] \\
	V_{i \mid i} &=\mathbb{E}\left[\left(\textbf{Z}_{i}-\hat{\textbf{z}}_{i \mid i}\right)\left(\textbf{Z}_{i}-\hat{\textbf{z}}_{i\mid i}\right)^{T} \mid \textbf{Y}_{0: i}\right]=\left(\mathbb{I}_r-K_{i} V J_{i|i-1} \right) V_{i \mid i-1},
\end{align*}
where
\begin{equation*}
	K_{i}  =	(VV_{i|i-1} J_{i|i-1})^T( VJ_{i|i-1} V_{i|i-1} J_{i|i-1}^TV^T+\Sigma)^{-1},
\end{equation*}
and where the various quantities predicted from data up to time $t_{i-1}$, which are formally defined in Appendix \ref{app:estep}, are dependent on a current estimate of parameters $\boldsymbol{{\beta}}^*$.
As for the moments of $G(\textbf{Z}_i)$, these are approximated by
\begin{align*}
	\mathbb{	E}\big[ G(\textbf{Z}_i)|\textbf{Y}_{0:i}\big]
	&\approx \textbf{g}_{i|i}+\frac{1}{2}\text{vect}(V_{i|i}H_{i|i}),\\
	\mathbb{	V}\big[ G(\textbf{Z}_i)|\textbf{Y}_{0:i}\big]
	&\approx J_{i|i}V_{i|i}J_{i|i}^T,\label{V_G_i}
\end{align*}
with 
\begin{equation*}
	\textbf{g}_{i|i}=G(\textbf{Z})\vert_{\hat{\textbf{z}}_{i|i}},  \hspace{1cm}
	J_{i|i}=\frac{\partial G(\textbf{Z})}{\partial \textbf{Z}} \vert_{\hat{\textbf{z}}_{i|i}}, \hspace{1cm} H_{i|i}=\frac{\partial^2 G(\textbf{Z})}{\partial \textbf{Z}^2} \vert_{\hat{\textbf{z}}_{i|i}}.
\end{equation*}
In particular, note how these moments depend on the moments of $\textbf{Z}_i$ derived above, i.e., $\hat{\textbf{z}}_{i \mid i}$ and $V_{i \mid i}$, so the latter are the main quantities that need to be calculated at the E-step. 

Algorithm \ref{alg:EKF} summarizes the calculations required for the Kalman filter at the E-step of the algorithm, based on a current estimate of parameters, $\boldsymbol{{\beta}}^*$.
 \begin{algorithm}[tb]
	\caption{Extended Kalman Filter (E-step)}\label{alg:EKF}
	\begin{algorithmic}
		\Require $\textbf{Y},\boldsymbol{{\beta}}^*,V$
		\For{$i=1,\dots,N$}
		\begin{enumerate}
			\item \textit{ Prediction step}
			
			\hspace{1cm} $\hat{\textbf{z}}_{i|i-1}={\boldsymbol{\mu}}_i $
			
			\hspace{1cm} $V_{i|i-1}={\text{diag}(\boldsymbol{\mu}}_i)$
			\item \textit{ Update step}
			
			\hspace{1cm}$\hat{\textbf{z}}_{i|i}=\hat{\textbf{z}}_{i\mid i-1}+K_{i}\bigg[\textbf{Y}_{i}-\textbf{Y}_{i-1}-V\big(\textbf{g}_{i|i-1}+\frac{1}{2}\text{vect}(V_{i|i-1}H_{i|i-1})\big)\bigg] $
			
			\hspace{1cm}$V_{i \mid i} =\left(\mathbb{I}-K_{i} V J_{i|i-1} \right) V_{i \mid i-1}$
			
			with
			
			\hspace{1cm}$	K_{i}  =		(VV_{i|i-1} J_{i|i-1})^T( VJ_{i|i-1}  V_{i|i-1} J_{i|i-1}^TV^T+\Sigma)^{-1}$
			
			\hspace{1cm}$\mu_{ij}= \exp({\beta}_j) \prod_{l=1}^p  {{Y}_{lt_{i-1}}\choose k_{lj}}(t_i-t_{i-1})\quad \quad j=1,\dots,r$
			
		\end{enumerate}
		\EndFor
	\end{algorithmic}
\end{algorithm}
The Kalman filter predictions of the latent states are used in the evaluation of 
$Q(\boldsymbol{{\beta}}|\boldsymbol{{\beta}}^*)=\mathbb{E}[\ell_{\textbf{Z}| \textbf{Y} }(\boldsymbol{{\beta}})|\textbf{Y}_{0:N}]+\mathbb{E}[\ell_{ \textbf{Y}|\textbf{Z}\textbf{Y}}|\textbf{Y}_{0:N}]$,
with
\begin{align}
	\mathbb{E}[\ell_{\textbf{Z}| \textbf{Y}}(\boldsymbol{{\beta}})|\textbf{Y}_{0:N}]
=&-\frac{1}{2}Nr\log(2\pi) -\frac{1}{2}\sum_{i=1}^{N}\bigg\{ \log(|\text{diag}({\boldsymbol{\mu}}_i)|) \notag \\
&+\hat{\textbf{z}}_{i|i}^T\text{diag}({\boldsymbol{\mu}}_i)^{-1}\hat{\textbf{z}}_{i|i} \notag\\
&+\text{Tr}\big[\text{diag}({\boldsymbol{\mu}}_i)^{-1}\cdot V_{i|i}\big]-2\hat{\textbf{z}}^T_{i|i}\cdot\boldsymbol{1}+\boldsymbol{\mu}_i^T\cdot\boldsymbol{1}  \bigg\},
\label{eq:Mstep}
\end{align}
\begin{align}
\mathbb{E}[\ell_{ \textbf{Y}|\textbf{Z}\textbf{Y}}|\textbf{Y}_{0:N}]= &-\frac{1}{2}N p \log \left(2 \pi \sigma^2\right)	\notag\\
		&-\frac{1}{2}\sigma^{-2}\sum_{i=1}^{N}\bigg\{\Delta \textbf{Y}_i^{T} \Delta \textbf{Y}_i -2 \Delta \textbf{Y}_i^{T}V \big(\textbf{g}_{i|i}+\frac{1}{2}\text{vect}(V_{i|i}H_{i|i})\big)\notag \\
		&+ (\textbf{g}_{i|i}+\frac{1}{2}\text{vect}(V_{i|i}H_{i|i})\big)^T V^{T} V  (\textbf{g}_{i|i}+\frac{1}{2}\text{vect}(V_{i|i}H_{i|i})\big) \notag\\
		&+\text{Tr}\left(V^{T} V J_{i|i}V_{i|i}J^{T}\right)\bigg\},
	\label{Qfun2}
\end{align}
and $\boldsymbol{{\beta}}^*$ the current value of the parameters used for the Kalman filter quantities $\hat{\textbf{z}}_{i|i}$ and $V_{i|i}$.

 \subsection{M-step}
The M-step maximizes the conditional expectation of the complete log-likelihood with respect to the parameters. Thus, the M-step involves the maximisation of the Q-function (\ref{Qfun}) with respect to $\boldsymbol{\beta}$.
Since the second term (\ref{Qfun2}) does not depend directly on $\boldsymbol{\beta}$, the M-step results in the optimization of the first term (\ref{eq:Mstep}). % where predictions of the latent variables and their covariance from the extended Kalman filtering at the E-step are based on the current estimate $\boldsymbol{{\beta}}^*$ of the parameters.

The optimal value of $\boldsymbol{\beta}$ from the M-step is  used as the new $\boldsymbol{\beta}^*$ for computing a new expected log-likelihood at the E-step. This iterative procedure is repeated until convergence, e.g., until the estimates of $\boldsymbol{\beta}$ do not change significantly. Algorithm \ref{alg:EM} summarizes the proposed EM algorithm.
 \begin{algorithm}
 	\caption{EM algorithm}\label{alg:EM}	
 	\begin{algorithmic}
 		\Require $\textbf{Y}, V, \boldsymbol{\beta}_{ini}, \sigma^2, tol, maxit$
 		\While{$err \geq tol$ $\&$ $it<maxit$}
 			\For{$i=1,\dots,N$}
        \begin{enumerate}
\item  E-step:
	
 \hspace{0.5cm} \textit{Extended Kalman Filter}: calculate $\hat{\textbf{z}}_{i|i}$ and $V_{i|i}$ from $\textbf{Y}$, $V$ and $\boldsymbol{\beta}_{old}$
 
\item M-step:
 		    \end{enumerate}
 	    
 \hspace{0.5cm}   $\boldsymbol{\beta}_{new}\gets\arg \max_{\boldsymbol{\beta}} Q(\boldsymbol{\beta}|\boldsymbol{\beta}_{old},\hat{\textbf{z}}_{i|i},V_{i|i})  $

 		 \hspace{0.5cm} $err \gets \max||\boldsymbol{\beta}_{new}-\boldsymbol{\beta}_{old}||_1^1$
 		 
 		 \hspace{0.5cm} $\boldsymbol{\beta}_{old}\gets\boldsymbol{\beta}_{new}$
 		 
 		  \hspace{0.5cm}  $it\gets it+1$.
		 \EndFor
 		\EndWhile
 
 	\end{algorithmic}
 \end{algorithm}

 %%%%%%%%%%%%%%%%%%%%%%%%%%%%%%%%%%%
 \subsection{Computational cost}
The computational cost of the proposed EM algorithm is the combination of the computational cost of the E- and M-steps. At the E-step, the latent variables $\textbf{Z}_i$ across the $N$ time intervals are of dimension $r$ and their covariance $V_{i|i}$ requires the inversion of a $p\times p$ matrix. Thus, the total complexity of the E-step is $\mathcal{O}(Nrp^3)$. On the other hand, the M-step concerns the optimization of an $r$-dimensional vector of parameters $\boldsymbol{\beta}$  and involves $N$ inversions of an $r\times r$ matrix for the calculation of the objective function. Thus, the total complexity of the M-step is $\mathcal{O}(Nr^3p)$. This results in a computational cost of the full algorithm of the order $\mathcal{O}(Nr^3p^3)$, although this may vary depending on the speed of convergence of the numerical algorithm used for the optimization of the Q-function at the M-step.  %Algorithm convergence strongly depends on the choice of the initial values, which in case of relatively small time intervals may be a problem.
  \subsection{Model selection}
In empirical settings, one may be interested in comparing different quasi-reaction systems of possibly varying complexity. 
 Since the marginal log-likelihood in (\ref{eq:marglik}) is not a direct result of the EM algorithm, we perform model selection by considering a modified version of the Bayesian Information Criterion, where the log-likelihood is replaced by the Q-function, which is instead a direct output of the EM algorithm. In particular, following \cite{ibrahim2008model}, we select the model that minimizes the following criterion
 \begin{equation}
 	BIC=-2Q(\boldsymbol{\hat{\beta}}|\boldsymbol{\hat{\beta}})+r\log(N).
 	\label{BIC}
 \end{equation}
where $r$ is the number of reaction rates in the model, $N$ the number of time intervals and $Q(\boldsymbol{\hat{\beta}}|\boldsymbol{\hat{\beta}})$ is
the Q-function (\ref{Qfun}) evaluated upon convergence of the EM algorithm.

\section{Simulation Study} \label{sec:5}
\begin{figure}[t!]
	\centering
	\begin{subfigure}{.35\textwidth}
		\centering
		\includegraphics[width=0.9\textwidth]{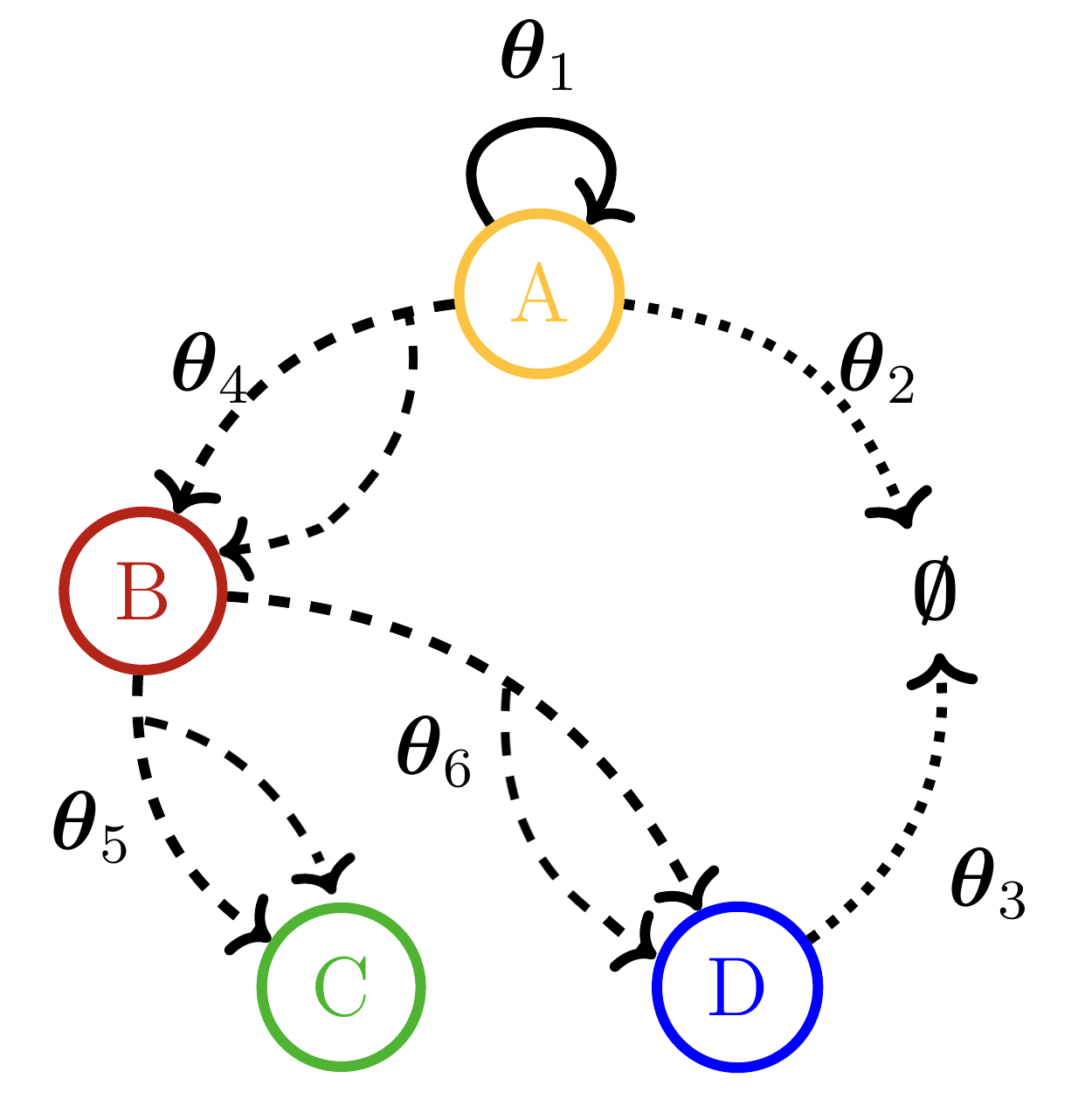}
 	 	\caption{}
\label{graph}
\end{subfigure}
		\begin{subfigure}{.2\textwidth}
			\centering
			\begin{align}
				\emptyset & \xrightarrow  {\theta_1} A \notag\\
				A & \xrightarrow{\theta_2}  \emptyset \notag\\
				D & \xrightarrow{\theta_3}  \emptyset \notag\\
					A & \xrightarrow{\theta_4}  2B \notag\\
				B & \xrightarrow{\theta_5}  2C \notag\\
				B & \xrightarrow{\theta_6}  2D \notag
			\end{align}
 	 	\caption{}
\label{eq}
\end{subfigure}
		\begin{subfigure}{.35\textwidth}
		\centering
		\includegraphics[width=1\textwidth]{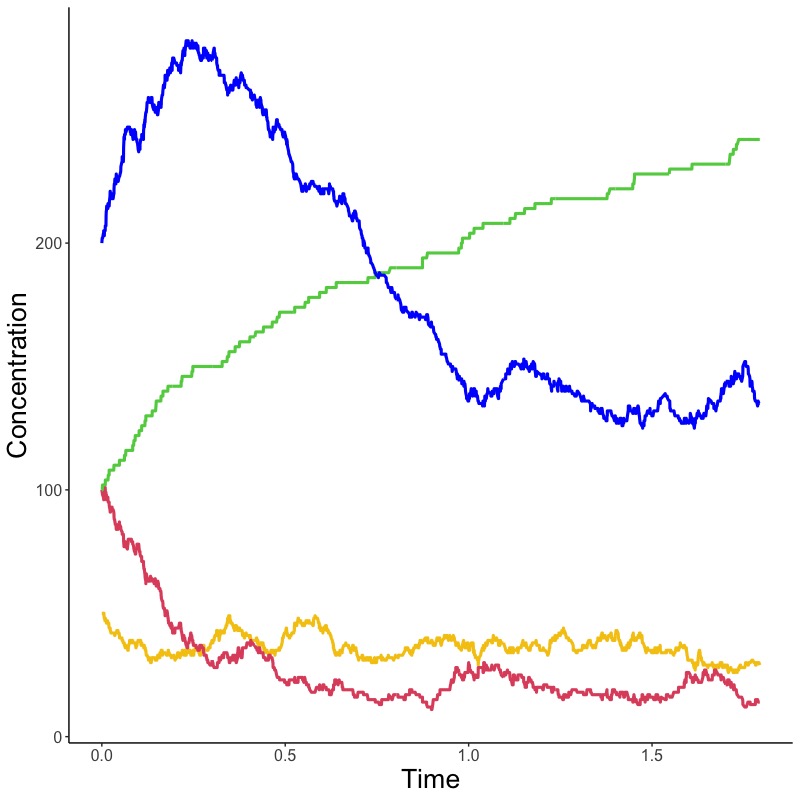}
 	 	\caption{}
\label{dyn}
\end{subfigure}
	\caption{\textbf{Specifications of the cell differentiation process used in the simulation study}. (a) Structure of the process with $p=4$ particles. Each substrate is represented by a coloured node, whereas birth, death and differentiation reactions are denoted by full, dotted and dashed edges, respectively. (b) The corresponding quasi-reaction system. (c) An example of trajectories generated by means of a Gillespie algorithm.}
\label{fig7}
\end{figure}

In this section, a simulation study is provided to evaluate the performance of the proposed method under different settings and to highlight those where it is particularly advantageous compared to the existing LLA approaches. For the simulation, we consider a dynamic system with a low number of particles ($p=4$) and reactions ($r=6$) in order to mimic a setting that is common in many applications, such as the cell differentiation process studied by \cite{pellin22}. In the specifics, the 6 reactions contain one duplication, two death and three differentiation reactions. Figure \ref{graph} provides a graphical representation of the system, while Figure \ref{eq} reports the 6 reactions. These corresponds to the net effect matrix
\[
V=\begin{pmatrix}
  1 & -1 & 0 & -1 & 0 & 0 \\
  0 & 0 & 0 & 2 & -1 & -1 \\
  0 & 0 & 0 & 0 & 2 & 0 \\
  0 & 0 & -1 & 0 & 0 & 2
\end{pmatrix}.
\]
As for the parameters, we set  
\[\boldsymbol{{\beta}}_{true}=\log(\boldsymbol\theta_{true})=( 5.30,  1.10, -0.11, -0.22, -0.22 ,-1.61)^T.\]
Starting with initial concentrations set to $\textbf{y}_0=(50,100,100,200)$, a Gillespie algorithm is used to generate the stochastic process over time \citep{Gillespie}.  Figure \ref{dyn} reports one run of the algorithm. %In the next section, we consider $100$ simulations for each scenario. 
 
  \paragraph{Improvement over local linear approximation approach}

In the first simulation study, we compare the performance of the algorithm against the existing LLA approach in terms of parameter estimation. Given the motivation behind the proposed methodology, we expect  an improvement when the interval between consecutive observations is particularly small, as this generates a strong temporal correlation among the particle concentrations. Moreover, we expect the difference to be more pronounced at low sample sizes, i.e., a small number of time points, as this will make statistical inference more challenging in general and may amplify the effect of strong temporal correlation.
	
In order to test these hypotheses, we consider a subset of the trajectories generated by the Gillespie algorithm. In particular, we consider the case where observations are retained at every 10, 15, 20, 25 and 30 time points out of the originally sampled trajectories. We refer to these as \textit{jumps}. The larger the jump is, the larger the gap between consecutive time points where the process is observed. This will generally translate into a large number of reactions that may have occurred between one time point and the next, although this will depend also on the dynamics of the process at the specific time interval. In order to test also the effect of sample size, in each of the settings, we consider two scenarios: one where we consider the first $N=5$ time intervals and a second one where we consider the first $N=50$ time intervals generated as above.
	
We perform parameter estimation for each of the datasets using LLA and our proposed EM algorithm. The LLA approach calculates the generalised least-square solution from the modelling assumption (\ref{gaussian}), given the concentration data $\textbf{Y}$. For the EM algorithm described in  Algorithm \ref{alg:EM}, we set the starting values $\boldsymbol{\beta}_{ini}$ to the LLA solution, $\sigma^2=0$, the net effect matrix to $V$ defined above, the tolerance for convergence to $tol=0.002$ and the maximum number of iterations to $maxit=300$. Upon convergence, we evaluate the quality of the estimation by calculating the Kullback-Leibler divergence between the estimated and the true parameters. In particular, this is defined by
  \begin{equation*}
	KL(\hat{\boldsymbol{\beta}},\boldsymbol{\beta}_{true})= \mathbb{	E}_{\textbf{y}^+}[\log p (\textbf{y}^+|\boldsymbol{\beta}_{true})-\log p(\textbf{y}^+|\hat{\boldsymbol{\beta}})],
\end{equation*}
where $\textbf{y}^+$  indicates an additional dataset with the same characteristics as the one used for inference, and generated from the same underling process defined by $\boldsymbol{{\beta}}_{true}$. The lower this value is, the closer the inferred process is to the true one.

\begin{figure}[tb!]
 	\centering
 	\begin{subfigure}[b]{0.45\textwidth}
 		\includegraphics[width=\textwidth]{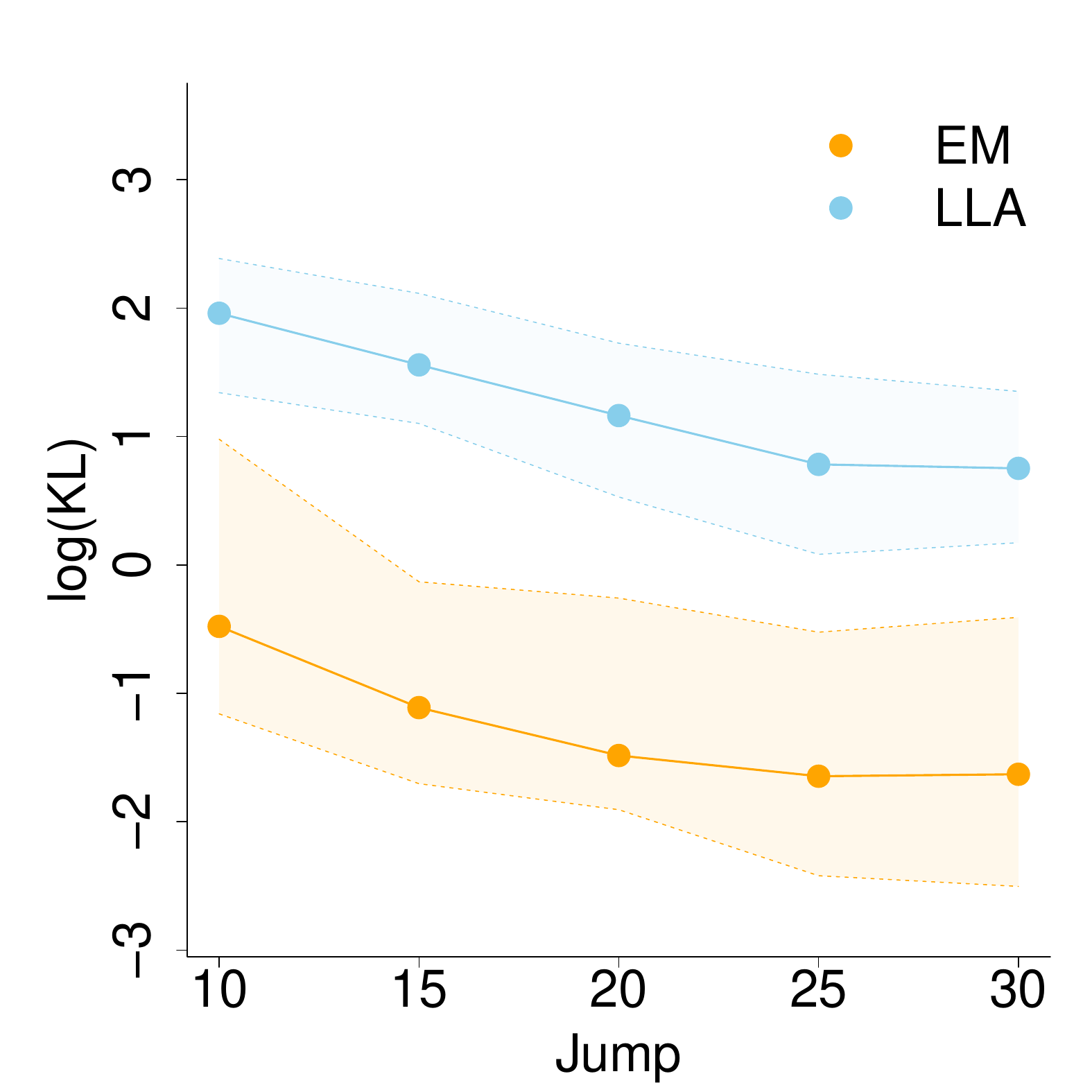}
 		\caption{}
 		\label{KL5}
 	\end{subfigure}
 	\begin{subfigure}[b]{0.45\textwidth}
 		\includegraphics[width=\textwidth]{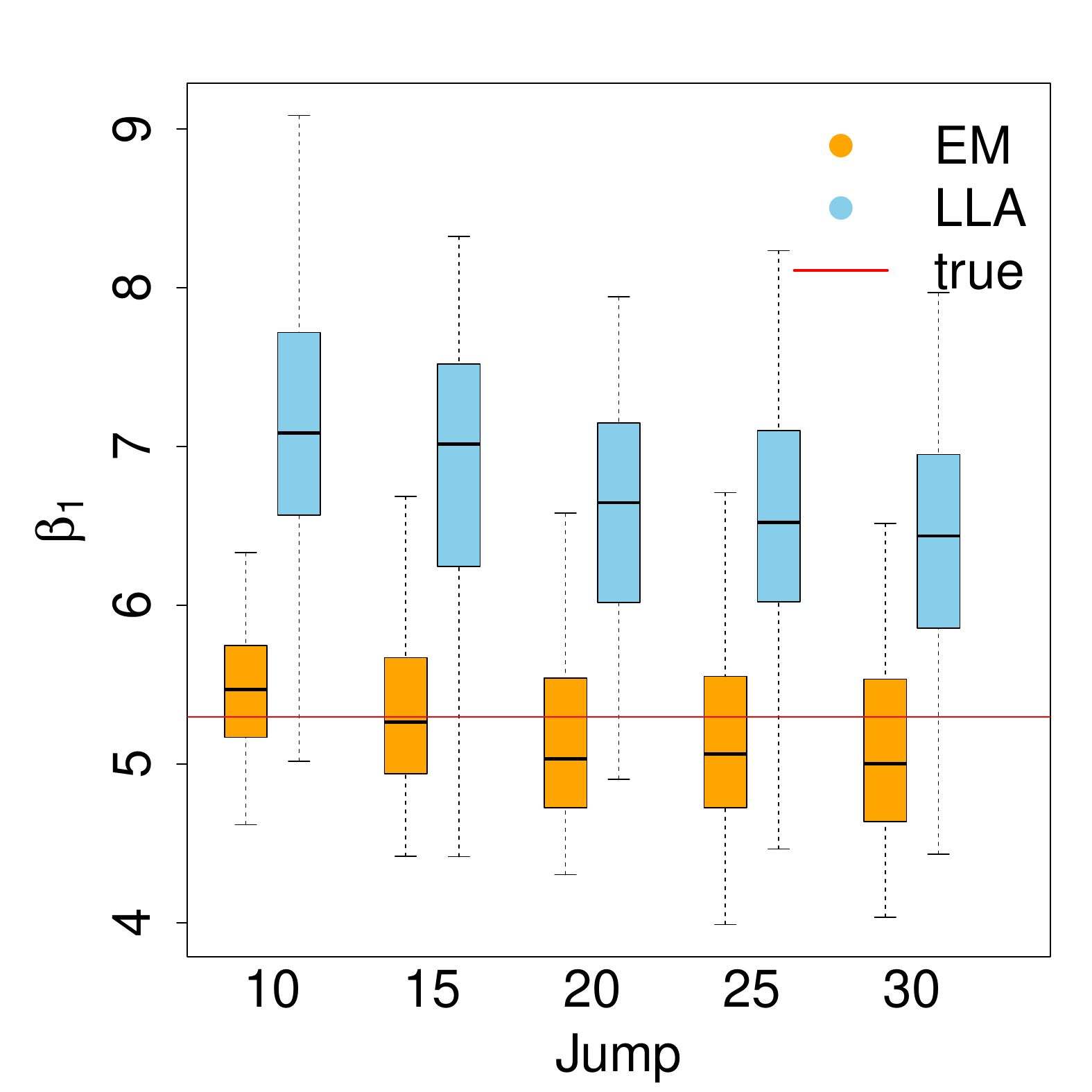}
 		\caption{}
 		\label{beta5}
 	\end{subfigure}
 	\begin{subfigure}[b]{0.45\textwidth}
 		\includegraphics[width=\textwidth]{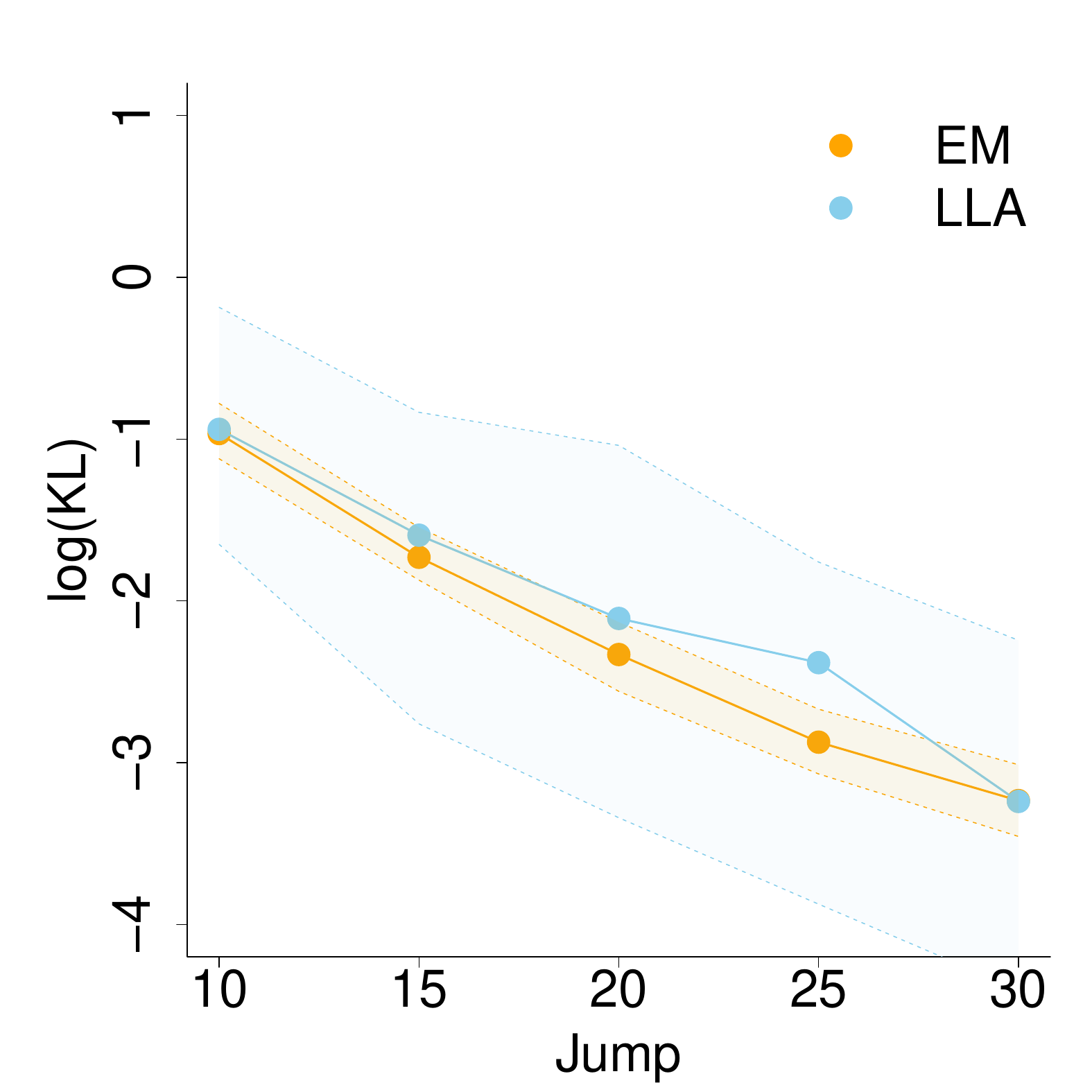}
 		\caption{}
 			\label{KL50}
 	\end{subfigure}
 	\begin{subfigure}[b]{0.45\textwidth}
 		\includegraphics[width=\textwidth]{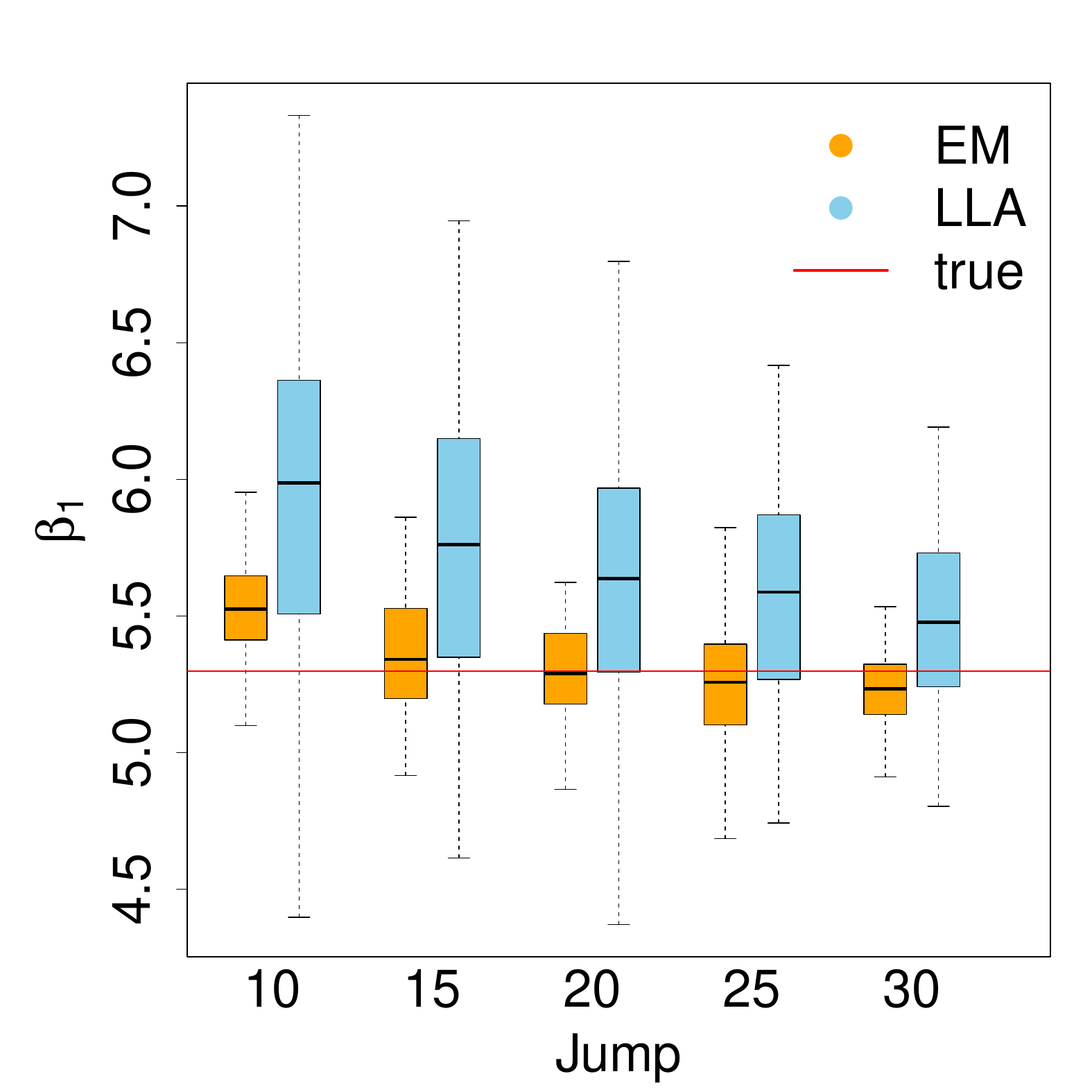}
 		\caption{}
 			\label{beta50}
 	\end{subfigure}
 	\caption{\textbf{Comparison between EM and LLA methods.} KL measure, in the log scale, shows that parameter estimation with the EM algorithm is closer to the true parameters than with the LLA approach.   On the right, the plots show how estimates of one of the parameters ($\beta_1$) are more accurate with the EM than with the LLA approach. All plots show how the effects are more pronounced with $N=5$ (\ref{KL5},\ref{beta5}) than with $N=50$ (\ref{KL50},\ref{beta50}) time intervals. The boxplots are obtained across 100 simulations.}
 	\label{compareJ50}
 \end{figure}
 
Figure \ref{compareJ50} reports the results in the form of boxplots across the 100 simulations for each of the settings.The results show how parameter estimation with the proposed EM algorithm is better than with the existing LLA approach, both in terms of the KL divergence (left panel) and estimation of one of the parameters ($\beta_1$, right panel).   All plots show how the effects are more pronounced with small sample sizes ($N=5$, Figure \ref{KL5} and \ref{beta5}) than with larger sample sizes ($N=50$, Figures \ref{KL50} and \ref{beta50}). Finally, Figure \ref{beta50} in particular shows how  the two approaches tend to converge to a similar performance for larger time intervals (i.e., a large \textit{jump}). This is to be expected, since temporal correlation will become less strong the larger the time interval. At the same time, the reconstruction of the reactions that have taken place within that time interval will also be less accurate. However,  Figure \ref{beta50} shows how, even in this case, estimation from the EM algorithm appears less biased and more accurate than with the LLA approach.

\begin{figure}[tb!]
	\centering
	\begin{subfigure}[b]{0.32\textwidth}
		\includegraphics[width=\textwidth]{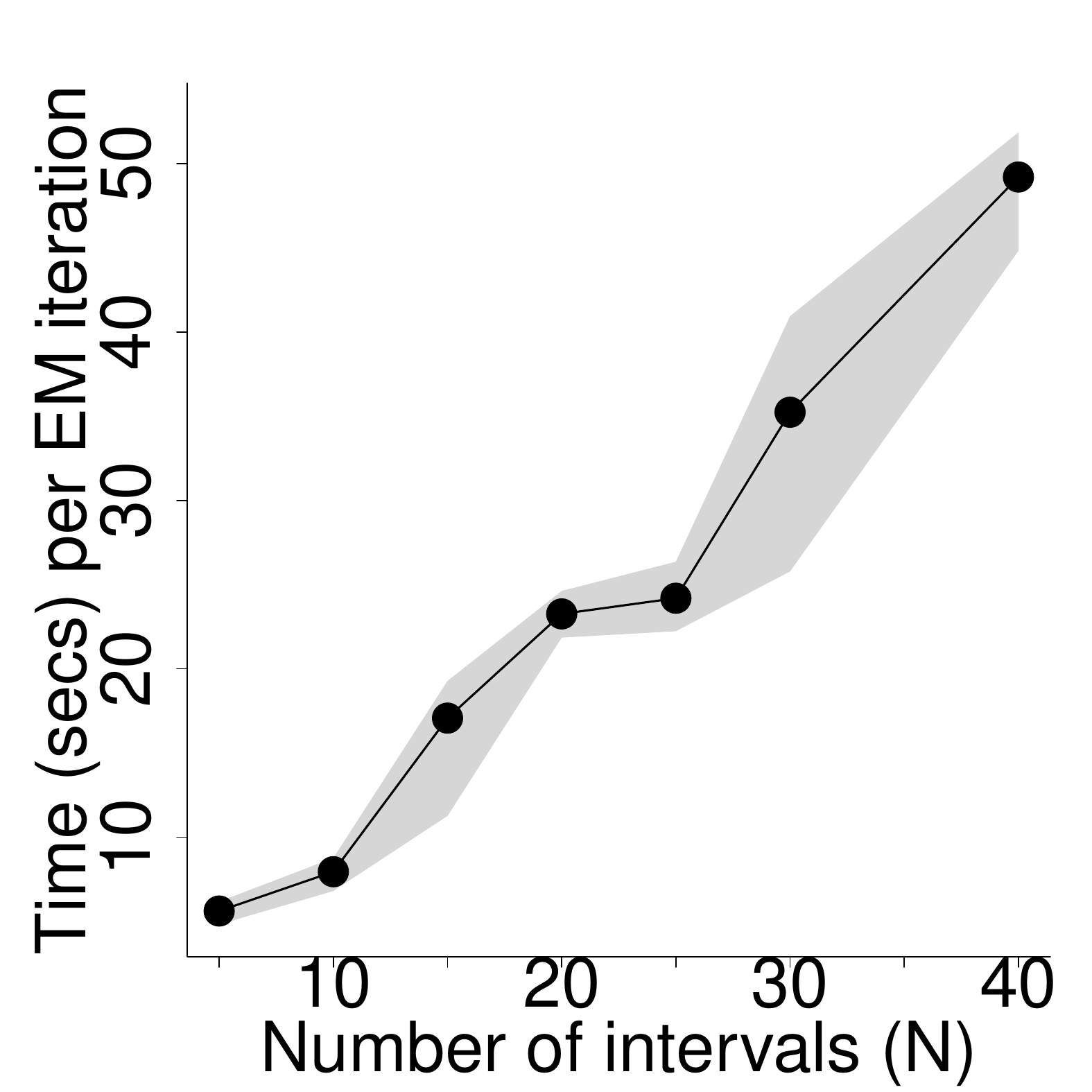}
		\caption{}
		\label{figobs}
	\end{subfigure}
	\hfill
	\begin{subfigure}[b]{0.32\textwidth}
		\includegraphics[width=\textwidth]{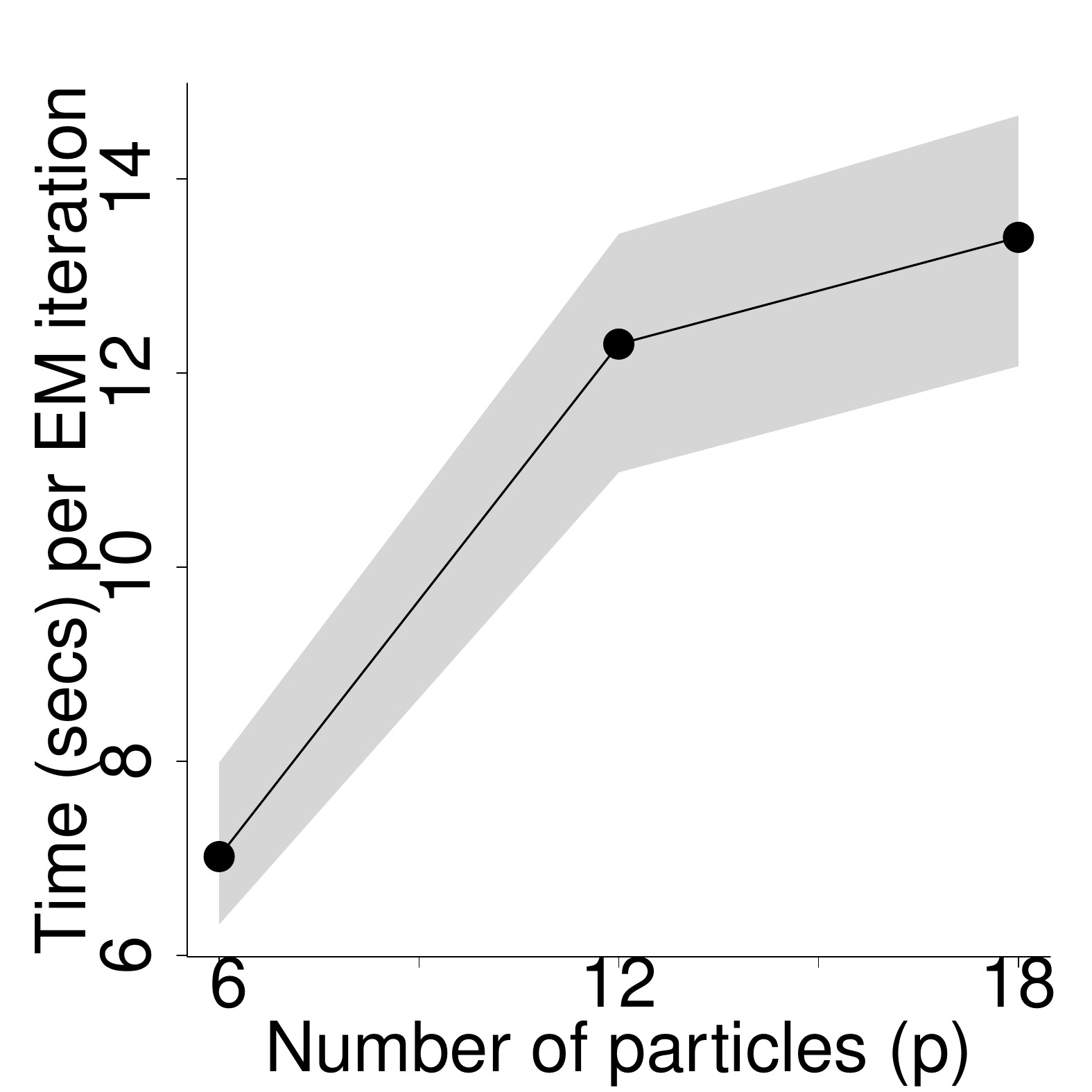}
		\caption{}
		\label{figP}
	\end{subfigure}
	\begin{subfigure}[b]{0.32\textwidth}
		\includegraphics[width=\textwidth]{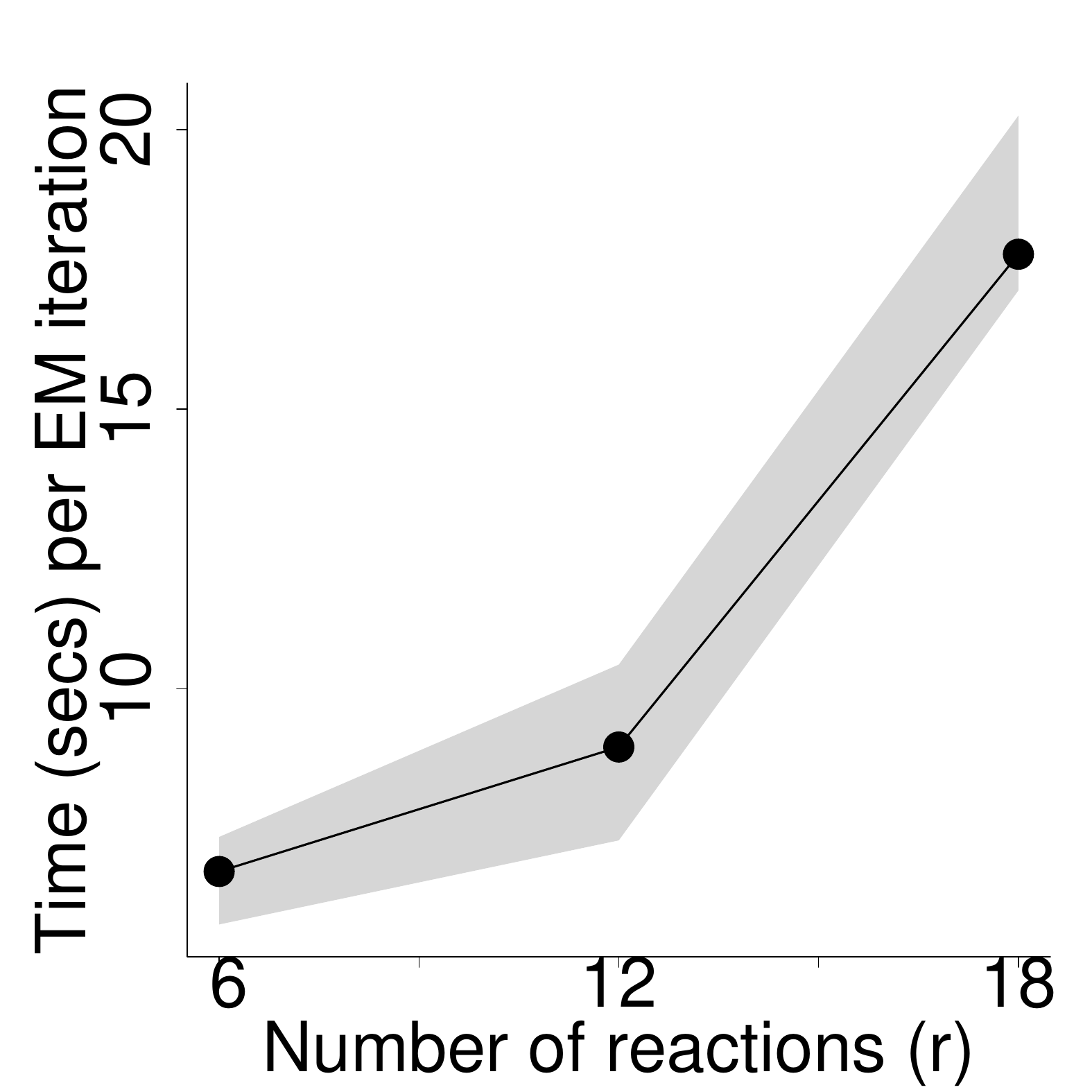}
		\caption{}
		\label{figR}
	\end{subfigure}
	\caption{\textbf{Computational cost of EM algorithm.} Average computational time (in seconds) of one iteration of the EM algorithm in terms of (a) the number of time intervals ($N$), (b) the number of particles ($p$), (c) the number of reactions ($r$). Median, first and third quartiles are shown across 100 simulations.}
	\label{fig:compareR}
\end{figure}

  \paragraph{Computational cost in terms of number of time points, reactions, particles}
In a second simulation study, we explore how the computational cost of the algorithm varies with respect to the number of time points ($N$), the number of reactions ($r$) and the number of particles ($p$). The results are shown in Figure \ref{fig:compareR}. In the first scenario (Figure \ref{figobs}), we consider the same generative process as before, fixing $jump=30$ and letting the number of time intervals vary in  $N=5,10,15,20,25,30,40$.   The plot shows how the average computational time of the EM algorithm is approximately linear in $N$. %The median maximum cost of the algorithm is $300N$ seconds.  

In the second scenario (Figure \ref{figP}), we evaluate how the computational time varies with respect to the number of particles $p$.  We set $jump=40$, $N=10$ and consider the three systems in Table \ref{var p} of Appendix \ref{app:dynsys}. These are characterized by the same number of reactions as before ($r=6$), but an increasing number of particles, namely $p=6, 12, 18$, respectively.  
 Figure \ref{figP} does not show the cubic dependence in $p$, that we had anticipated. Given that the computational time is the combined time from the E- and the M-steps, this suggests a much slower M-step.

Finally, in the third scenario (Figure \ref{figR}), we evaluate how the computational time varies with respect to the number of reactions $r$. As before we set $jump=40$ and $N=10$, but we now consider the three systems in Table \ref{var r} of Appendix \ref{app:dynsys}. These are characterized by the same number of particles as before ($p=6$), but an increasing number of reactions, namely $r=6, 12, 18$, respectively. The plot shows a super-linear dependence in $r$.

\section{Illustration on Italian COVID-19  data}\label{sec:6}
%In epidemiology and public health research, it is common to work with correlated data, as the outcome of interest is often clustered in groups.
We provide a real data illustration focusing on the COVID-19 pandemic. Worldwide, more than 700 million infections and almost 7 million deaths were recorded as of August 16, 2023 \citep{covid2020dashboard}. As a result, significant efforts have been made in order to understand the phenomenon and  find strategies to slow down the spreading of the disease. Italy has been one of the countries of interest during the pandemic, being the first European country to experience a significant outbreak of the disease \citep{liao2020tw}. The first case was confirmed on $31$ January $2020$. Since then, for more than two years, data were collected daily. The sufficiently close interval between observations is ideal for the application of our method, as it generates strong temporal correlations. In the following analysis, we focus on daily data within three specific time intervals, characterized by three different levels of contagion:
\begin{itemize}
	\item \textbf{Phase 1}: \textit{$9^{\text{th}}$March - $4^{\text{th}}$May, $2020$}. Strong restrictions on travel throughout the country, banning all forms of gathering in private and public places \citep{conte1};
	\item \textbf{Phase 2}: \textit{$4^{\text{th}}$May - $7^{\text{th}}$October, $2020$}. Containment measures were relaxed, allowing the travelling for visits to relatives (within a region) and the restart of several production activities \citep{conte2};
	\item \textbf{Phase 3}: \textit{ $8^{\text{th}}$October, $2020$ - $14^{\text{th}}$January, $2021$}. Wearing of masks became compulsory both outdoor and indoor, and assemblages were restricted \citep{conte3}.
\end{itemize}

Within each of the three phases and using data from all 21 Italian regions, we use the proposed EM algorithm to fit the parameters of the following two dynamic systems:
			\begin{align}
   		\textbf{Model A} & \quad &\textbf{Model B} \notag \\
			I_k  \xrightarrow  {\theta_{1k}} 2I_k & \quad & I_k  \xrightarrow  {\theta_{1k}} 2I_k\notag\\
			I_k  \xrightarrow{\theta_{2}}  R_k & \quad & I_k  \xrightarrow{\theta_{2k}}  R_k \notag\\
			I_k \xrightarrow{\theta_{3}}  D_k & \quad & I_k \xrightarrow{\theta_{3k}}  D_k \notag
		\end{align}
That is, we fit a simple SIR model where the number of susceptible individuals \textit{S} has been omitted, as it is almost constant throughout the observation period, \textit{I} is the number of infectious individuals, \textit{R} the number of recovered individuals and \textit{D} the number of deceased individuals \citep{Simon}. The index $k$ denotes the region. Thus, model B is characterized by region-specific rates for all three reactions. This results in a system with $p=63$ particles and $r=63$ rates.  On the other hand, model A hypothesizes a simpler model where the recovery and death rates are assumed to be the same across Italy, under an assumption that these depend primarily on the specifics of the virus and are not as affected by the level of contagion in the population. 

 For the EM algorithm, we set the parameters to $\sigma^2=0$ and $tol=0.002$. Using Equation (\ref{BIC}), model A and model B result in $2.42\cdot 10^6$ and $1.92\cdot 10^6$ BIC values, respectively. This leads to the choice of the more complex model B, with region-specific recovery and death reaction rates. 
\begin{figure}[t!]
	\centering
	\includegraphics[width=1\textwidth]{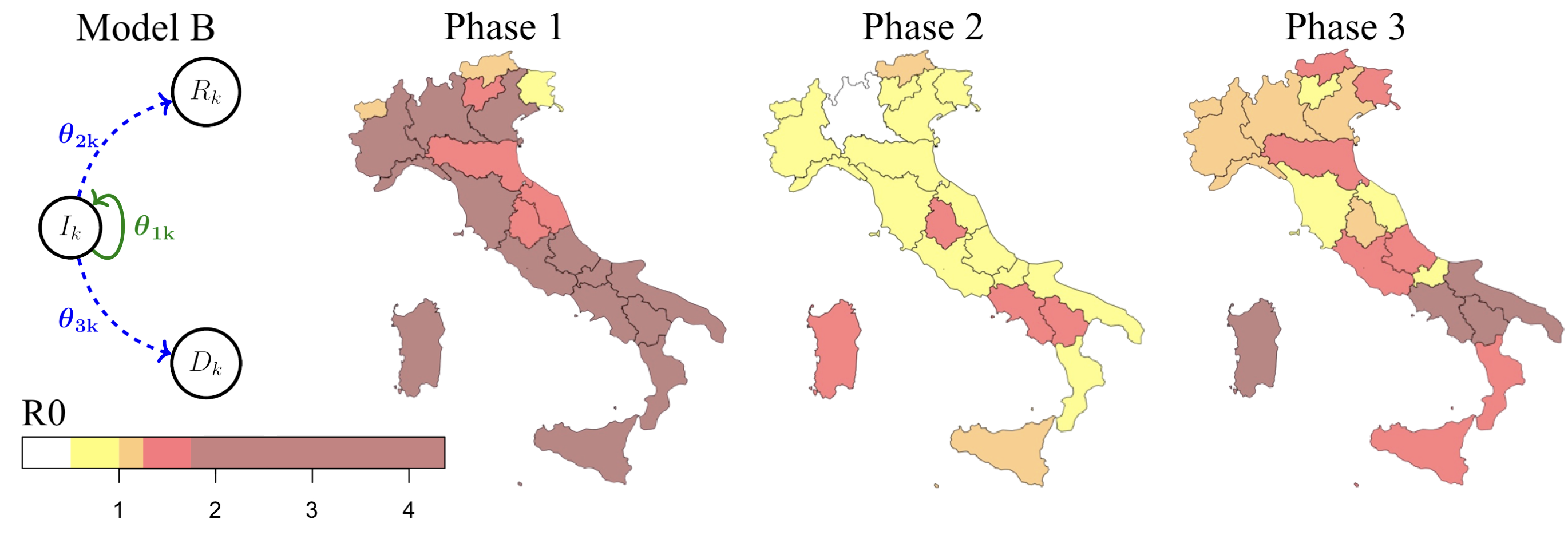}
	\caption{\textbf{Visualization of the estimated $R0$ values in Italy.}  On the left, the  system of kinetic reactions for the $k$-th region. On the right, a visualization of the estimated values of the basic reproductive numbers for each Italian region, in the 3 phases of interest, categorised according to the colour coding used by the Italian government.}
\label{fig:covid}
\end{figure}
Figure \ref{fig:covid} visualises the results in terms of the estimated basic reproduction number $R0$, i.e., $\theta_{1k}/(\theta_{2k} +\theta_{3k})$. These values are categorized according to the colour coding used by the Italian government to evaluate the severity of the disease spread. %; we have added a darker shade of red to better appreciate the results in Phase $1$ and Phase $3$. 
During Phase 2 the infection was limited, which was also a consequence of the containment measures implemented in Phase 1. During Phase 3, a revival of the disease spread is observed, in particular in the southern regions of Italy. The estimated $R0$ values are in line with those from other studies \citep{giordano2020modelling, remuzzi2020covid,mingliang2022r}.

\section{Conclusions}\label{sec:conclusion}
In this paper, we have proposed a novel procedure for the statistical inference of quasi-reaction systems.
 Local linear approximation methods tend to perform poorly when the system is observed at fine time intervals. This is due to numerical instability caused by strong correlations in the observations from one time point to the next.  The proposed method focuses instead on reconstructing the underlying process of latent reactions.
To this end, we develop a latent event history model of the observed count process driven by a latent process of reactions. We propose a computationally efficient EM algorithm for parameter estimation, with an extended Kalman filtering procedure for the prediction of the latent states. A
simulation study shows how the proposed method performs better than the existing LLA approach, particularly when the time intervals between consecutive observations are small and the number of time points is low.

The method is illustrated by an application on the Italian Covid 19 data. We analyse daily data provided by the 21 Italian regions from March 2020 to January 2021, the critical phase of the pandemic. The basic reproduction number $R0$ of the 21 Italian zones estimated by our method in three consecutive phases of the pandemic shows higher values at the beginning and at the end of the time period. This is to be expected given the evolution of the disease and the societal restrictions that were imposed by the Italian government during this period. %certified by actual events and the state of the art. %From a secondary analysis, a model selection is performed:  a scheme was preferred in which the parameters were not common among the regions, certifying how the recovery and death rates were dependent on regional factors. 

\section*{Data availability}
The data used in this paper are available from the Dipartimento della Protezione Civile and the Istituto Nazionale di Statistica (ISTAT) via the webpage of the department (\url{http://dati.istat.it/Index.aspx?QueryId=18460}) and a Github repository (\url{https://github.com/pcm-dpc/COVID-19}).

\appendix
\section{Kalman filtering (E-step)} \label{app:estep}
In this section, we discuss the extended Kalman filtering procedure that was developed for the evaluation of $\mathbb{E}[\textbf{Z}_i|\textbf{Y}_{0:i}] $, $\mathbb{V}[\textbf{Z}_i|\textbf{Y}_{0:i}]$, $\mathbb{E}[G(\textbf{Z}_i)|\textbf{Y}_{0:i}] $, and $\mathbb{V}[G(\textbf{Z}_i)|\textbf{Y}_{0:i}]$.

\paragraph{Prediction Step}
The prediction step calculates the first and second moments of $\textbf{Z}_i$ conditional on $\textbf{Y}_{0:i-1}$. According to model (\ref{statespace2}), these are in fact the conditional moments of $\textbf{Z}_i$ given the state of the system at the previous time point,  $\textbf{Y}_{i-1}$. Thus 
\begin{align}
	\hat{\textbf{z}}_{i\mid i-1}&=\mathbb{E}\left[\textbf{Z}_{i}\mid \textbf{Y}_{i-1}\right]=\boldsymbol{\mu}_i \notag\\ 
	V_{i \mid i-1}&=\mathbb{V}\left[\textbf{Z}_{i}\mid \textbf{Y}_{i-1}\right]=\text{diag}(\boldsymbol{\mu}_i).
	\label{eq:predict}
\end{align}

\paragraph{Update Step}
Following from the prediction step, the update step refines these predictions by comparing them to the observed values at time $i$. In particular, we update the conditional distribution of $\textbf{Z}_i$ from the past with the information coming from $\textbf{Y}_i$ by first deriving the joint distribution of $\textbf{Y}_i$ and $\textbf{Z}_i$ conditional on $\textbf{Y}_{0:i-1}$. According to model (\ref{statespace2}), this is a multivariate Gaussian distribution, which we write  generically as
\begin{equation}
	\begin{array}{l}
		\textbf{Z}_i \\
		\textbf{Y}_i
	\end{array} \bigg|	\textbf{Y}_{0:i-1} \sim \mathcal{N}\left(\left[\begin{array}{c}
	\textbf{m}_1\\
		\textbf{m}_2
	\end{array}\right],\left[\begin{array}{cc}
	S_{11} &S_{12} \\
	S_{21} & S_{22}
	\end{array}\right]\right).
\label{eq:updatedist}
\end{equation}
From the prediction step (\ref{eq:predict}), we already know $\textbf{m}_1$ and $S_{11}$. 
As regards to the other elements of the mean and covariance, we have
\begin{align}
	\textbf{m}_2&= 	\mathbb{E}[\textbf{Y}_i|\textbf{Y}_{0:i-1}]=\mathbb{E}[\textbf{Y}_{i-1}+VG(\textbf{Z}_i)+\boldsymbol{\psi}_i|\textbf{Y}_{0:i-1}] \notag \\
	&=\textbf{Y}_{i-1}+V\mathbb{E}[G(\textbf{Z}_i)|\textbf{Y}_{0:i-1}] \notag\\
S_{12}&=  \mathbb{C}ov[\textbf{Z}_i,\textbf{Y}_i|\textbf{Y}_{0:i-1}] \notag\\
&= \mathbb{C}ov[\textbf{Z}_i,\textbf{Y}_{i-1}+VG(\textbf{Z}_{i})+\boldsymbol{\psi}_i|\textbf{Y}_{0:i-1}]= V\mathbb{C}ov[\textbf{Z}_i,G(\textbf{Z}_{i})|\textbf{Y}_{0:i-1}]V^T \notag\\
S_{22}&=	\mathbb{V}[\textbf{Y}_i|\textbf{Y}_{0:i-1}] =\mathbb{V}[\textbf{Y}_{i-1}+VG(\textbf{Z}_i)+\boldsymbol{\psi}_i|\textbf{Y}_{0:i-1}] \notag\\
&=V\mathbb{V}\big[G(\textbf{Z}_i)|\textbf{Y}_{0:i-1}\big]V^T+\Sigma.
\label{eq:update}
\end{align}

In order to calculate the first and second moments of $G(\textbf{Z}_i)$, we  approximate the non-linear function $G$ with its Taylor expansion of order $2$ centered  at $\hat{\textbf{z}}_{i|i-1}$, i.e.,
\begin{equation*}
	G(\textbf{Z}_i)\approx \textbf{g}_{i|i-1}+J_{i|i-1}(\textbf{Z}_i-\hat{\textbf{z}}_{i|i-1})+\frac{1}{2}\text{diag}(\textbf{Z}_i-\hat{\textbf{z}}_{i|i-1})H_{i|i-1}(\textbf{Z}_i-\hat{\textbf{z}}_{i|i-1}).
\end{equation*}
The first term of the expansion is the deterministic vector of size $r$
\begin{equation*}
	\textbf{g}_{i|i-1}=G(\hat{\textbf{z}}_{i|i-1}).
\end{equation*}
The other terms have a simplified form due to the fact that the $j$-th element of the function $G$ is function only of the $j$-th element of $\textbf{Z}_i$. Thus, the $r\times r$ matrix of first derivatives is  a diagonal matrix, with $(j,j)$ element given by
\begin{align*}
	\big[J_{i|i-1}\big]_{jj}&=\big[\frac{\partial G(\textbf{Z})}{\partial \textbf{Z}} \vert_{\hat{\textbf{z}}_{i|i}}\big]_{jj}=\frac{\partial G_j}{\partial z_{ij}}\bigg|_{\hat{z}_{ij|i-1}} \\
	&=\frac{\partial (F_{ij}^{-1}(\Phi_{ij}(Z_{ij}))}{\partial z_{ij}}\bigg|_{\hat{z}_{ij|i-1}}=\bigg( \frac{\partial F_{ij}}{\partial x_{ij}}\bigg|_{G(\hat{z}_{ij|i-1})}  \bigg)^{-1}\frac{\partial \Phi_{ij}}{\partial z_{ij}}\bigg|_{\hat{z}_{ij|i-1}},
\end{align*}
where, using the functional form of the Normal and Gamma CDFs, we have
\begin{equation*}
		\frac{\partial \Phi_{ij}}{\partial z_{ij}}(z) = \frac{e^{-\frac{(z- \mathbb{E}[Z_{ij}])^2}{2\mathbb{	V}[Z_{ij}]}}}{\sqrt{2 \pi\mathbb{	V}[Z_{ij}]} }, \hspace{3cm} 
	\frac{\partial F_{ij}}{\partial x_{ij}}(x)= \frac{e^{-x}x^{\mathbb{	E}[X_{ij}]-1}}{\Gamma(\mathbb{E}[X_{ij}])}\mathbbm{1}_{[x>0]}.
\end{equation*}
Similarly, the $r \times r \times r$ Hessian matrix, can be written as an $r \times r$ diagonal matrix with second derivatives on the diagonal, namely
\begin{align*}
	\big[	H_{i|i-1}\big]_{jj}&=\frac{\partial}{\partial z_{ij}}\bigg[\bigg(\frac{\partial F_{ij}}{\partial x_{ij}}\big|_{G(\hat{z}_{ij|i-1})}\bigg)^{-1}\frac{\partial \Phi_{ij}}{\partial z_{ij}}\bigg|_{\hat{z}_{ij|i-1}}\bigg]\\
		&=\frac{-\frac{\partial^2 F_{ij}}{(\partial x_{ij})^2}\bigg|_{G(\hat{z}_{ij|i-1})}\frac{\partial G_{j}}{\partial z_{ij}}\bigg|_{\hat{z}_{ij|i-1}}\frac{\Phi_{ij}}{\partial z_{ij}}\bigg|_{\hat{z}_{ij|i-1}}+\frac{\partial F_{ij}}{\partial x_{ij}}\bigg|_{G(\hat{z}_{ij|i-1})}\frac{\partial^2\Phi_{ij}}{(\partial z_{ij})^2}\bigg|_{\hat{z}_{ij|i-1}}}{\bigg(\frac{\partial F_{ij}}{\partial x_{ij}}\bigg|_{G(\hat{z}_{ij|i-1}))},\bigg)^2}
\end{align*}
where
\begin{equation*}
	\frac{\partial^2 \Phi_{ij}}{\partial z_{ij}^2}(z)  = \frac{(z-\mathbb{	E}[Z_{ij}])e^{-\frac{(z - \mathbb{	E}[Z_{ij}])^2}{2\mathbb{	V}[Z_{ij}]}}}{\sqrt{2 \pi} \mathbb{	V}[Z_{ij}]^{3/2}}, 
	\frac{\partial^2 F_{ij}}{\partial x_{ij}^2}(x) = \frac{e^{-x}x^{\mathbb{	E}[X_{ij}]-2}(\mathbb{	E}[X_{ij}]-x-1)}{\Gamma(\mathbb{E}[X_{ij}])}\mathbbm{1}_{[x>0]}.
\end{equation*}

Going back to (\ref{eq:update}), we can now use the Taylor approximation to calculate the required conditional expectations. In particular, we have
\begin{align*}
\mathbb{E}\big[ G(\textbf{Z}_i)|\textbf{Y}_{0:i-1}\big]
	\approx &\mathbb{E}\bigg[\textbf{g}_{i|i-1}+J_{i|i-1}(\textbf{Z}_i-\hat{\textbf{z}}_{i|i-1}) \notag \\
	& +\frac{1}{2}\text{diag}(\textbf{Z}_i-\hat{\textbf{z}}_{i|i-1}) H_{i|i-1}(\textbf{Z}_i-\hat{\textbf{z}}_{i|i-1})|\textbf{Y}_{0:i-1}\bigg]\notag\\
	= & \textbf{g}_{i|i-1}+J_{i|i-1}\mathbb{E}\bigg[\textbf{Z}_i-\hat{\textbf{z}}_{i|i-1}|\textbf{Y}_{0:i-1}\bigg] \notag \\
	& +\frac{1}{2}\mathbb{E}\bigg[\text{diag}(\textbf{Z}_i-\hat{\textbf{z}}_{i|i-1}) H_{i|i-1}(\textbf{Z}_i-\hat{\textbf{z}}_{i|i-1})|\textbf{Y}_{0:i-1}\bigg]\notag\\
	= & \textbf{g}_{i|i-1}+\frac{1}{2}\text{vect}(V_{i|i-1}H_{i|i-1}),\notag
	\end{align*}
	\begin{align*}
	\mathbb{C}ov[\textbf{Z}_i,G(\textbf{Z}_i)|\textbf{Y}_{0:i-1}]
	\approx & \mathbb{C}ov\bigg[\textbf{Z}_i,\textbf{g}_{i|i-1}|\textbf{Y}_{0:i-1}\bigg] \notag \\
	& +\mathbb{C}ov\bigg[\textbf{Z}_i,J_{i|i-1}(\textbf{Z}_i-\hat{\textbf{z}}_{i|i-1})|\textbf{Y}_{0:i-1}\bigg]	\notag\\
	&+\mathbb{C}ov\bigg[\textbf{Z}_i,\frac{1}{2}\text{diag}(\textbf{Z}_i-\hat{\textbf{z}}_{i|i-1}) H_{i|i-1}(\textbf{Z}_i-\hat{\textbf{z}}_{i|i-1})|\textbf{Y}_{0:i-1}\bigg]\notag\\
	= & \mathbb{C}ov\bigg[\textbf{Z}_i,J_{i|i-1}(\textbf{Z}_i-\hat{\textbf{z}}_{i|i-1})|\textbf{Y}_{0:i-1}\bigg] \notag \\
	 = & \mathbb{V}ar[\textbf{Z}_i|\textbf{Y}_{0:i-1}] J_{i|i-1}= V_{i|i-1} J_{i|i-1},
	\end{align*}
	\begin{align*}
		\mathbb{	V}\big[ G(\textbf{Z}_i)|\textbf{Y}_{0:i-1}\big]&\approx\mathbb{V}\bigg[\textbf{g}_{i|i-1}+J_{i|i-1}(\textbf{Z}_i-\hat{\textbf{z}}_{i|i-1})|\textbf{Y}_{0:i-1}\bigg]\\
		&= J_{i|i-1}\mathbb{V}\bigg[\textbf{Z}_i-\hat{\textbf{z}}_{i|i-1}|\textbf{Y}_{0:i-1}\bigg]J_{i|i-1}^T = J_{i|i-1}V_{i|i-1}J_{i|i-1}^T.
\end{align*}

Finally, plugging these expressions into (\ref{eq:predict}), we have
\begin{align*}
	\textbf{m}_2
	&\approx \textbf{Y}_{i-1}+V\bigg[  \textbf{g}_{i|i-1}+\frac{1}{2}\text{vect}(V_{i|i-1}H_{i|i-1})  \bigg]\notag\\
	S_{22}&\approx V[J_{i|i-1} V_{i|i-1} J_{i|i-1}^T]V^T+\Sigma\notag\\
	S_{12}&\approx VV_{i|i-1} J_{i|i-1},\notag
\end{align*}
which, together with $\textbf{m}_1$ and $S_{11}$ derived previously, define the joint distribution (\ref{eq:updatedist}) of $\textbf{Z}_{i}$ and $\textbf{Y}_{i}$ conditional on $\textbf{Y}_{i-1}$. From this, using the formulae for the  conditional distributions from a jointly Gaussian random vector, we conclude that $\textbf{Z}_{i}$ conditional on $\textbf{Y}_{0:i}$ has a multivariate Gaussian distribution, with mean and covariance given, respectively, by
\begin{align*}
\hat{\textbf{z}}_{i \mid i} &=\mathbb{E}\left[\textbf{Z}_{i} \mid \textbf{Y}_{0: i}\right]=\hat{\textbf{z}}_{i\mid i-1}+K_{i}\bigg[\textbf{Y}_{i}-\textbf{Y}_{i-1}-V\big(\textbf{g}_{i|i-1}+\frac{1}{2}\text{vect}(V_{i|i-1}H_{i|i-1})\big)\bigg] \\
	V_{i \mid i} &=\mathbb{E}\left[\left(\textbf{Z}_{i}-\hat{\textbf{z}}_{i \mid i}\right)\left(\textbf{Z}_{i}-\hat{\textbf{z}}_{i\mid i}\right)^{T} \mid \textbf{Y}_{0: i}\right]=\left(\mathbb{I}_r-K_{i} V J_{i|i-1} \right) V_{i \mid i-1},
\end{align*}
where
\begin{equation*}
	K_{i}  =	(VV_{i|i-1} J_{i|i-1})^T( VJ_{i|i-1} V_{i|i-1} J_{i|i-1}^TV^T+\Sigma)^{-1}.
\end{equation*}
Note how the update step refines the conditional expectation found in the prediction step in proportion to the difference between the actual and estimated observations, i.e., the prediction error. Moreover, this is directly proportional to the magnitude of the Kalman \textit{gain matrix} $K_i$,  which captures the linear relationship between the noise and the variance of the latent variable \citep{Kim}.

Similarly to the earlier derivations, we have that
\begin{align*}
	\mathbb{	E}\big[ G(\textbf{Z}_i)|\textbf{Y}_{0:i}\big]
	&\approx \textbf{g}_{i|i}+\frac{1}{2}\text{vect}(V_{i|i}H_{i|i}),\\
	\mathbb{	V}\big[ G(\textbf{Z}_i)|\textbf{Y}_{0:i}\big]
	&\approx J_{i|i}V_{i|i}J_{i|i}^T,\label{V_G_i}
\end{align*}
with 
\begin{equation*}
	\textbf{g}_{i|i}=G\vert_{\hat{\textbf{z}}_{i|i}},  \hspace{1cm}
	J_{i|i}=\frac{\partial G(\textbf{Z})}{\partial \textbf{Z}} \vert_{\hat{\textbf{z}}_{i|i}}, \hspace{1cm} H_{i|i}=\frac{\partial^2 G(\textbf{Z})}{\partial \textbf{Z}^2} \vert_{\hat{\textbf{z}}_{i|i}}.
\end{equation*}

\section{Dynamic systems used for the simulation study} \label{app:dynsys}
In this section, we report the systems of reactions that were used in Section \ref{sec:5} for evaluating the computational complexity of the algorithm with respect to the number of particles $p$ (Table \ref{var p}) and the number of reactions $r$ (Table \ref{var r}).
\begin{table}[h!]
	\tiny
	\centering
	\begin{minipage}{.19\textwidth}
		\centering
		$p=6$
		\begin{align}
			\begin{cases}
				Y_1 & \xrightarrow  {\theta_1} Y_2 \notag\\
				Y_1 & \xrightarrow{\theta_2} 	Y_3  \notag\\
				Y_2 & \xrightarrow{\theta_3} 	Y_4  \notag\\
				Y_1 & \xrightarrow{\theta_4}  	Y_4 \notag\\
				Y_4& \xrightarrow{\theta_5}  	Y_6 \notag\\
				Y_5 & \xrightarrow{\theta_6}  	Y_5 \notag
			\end{cases}
		\end{align}
	\end{minipage}
	\begin{minipage}{.4\textwidth}
		\centering
		$p=12$
		\begin{align}
			\begin{cases}
				Y_1+Y_7 & \xrightarrow  {\theta_1} Y_2+Y_8 \notag\\
				Y_1 +Y_7& \xrightarrow{\theta_2} 	Y_3 +Y_9 \notag\\
				Y_2 +Y_8& \xrightarrow{\theta_3} 	Y_4+Y_{10}  \notag\\
				Y_1+Y_7 & \xrightarrow{\theta_4}  	Y_4 +Y_{10}\notag\\
				Y_4+Y_{10}& \xrightarrow{\theta_5}  	Y_6+Y_{12} \notag\\
				Y_5+Y_{11} & \xrightarrow{\theta_6}  	Y_5+Y_{11} \notag
			\end{cases}
		\end{align}
	\end{minipage}
	\begin{minipage}{.4\textwidth}
		\centering
		$p=18$
		\begin{align}
			\begin{cases}
				Y_1+Y_7+Y_{13} & \xrightarrow  {\theta_1} Y_2+Y_8+Y_{14} \notag\\
				Y_1 +Y_7+Y_{13}& \xrightarrow{\theta_2} 	Y_3 +Y_9+Y_{15} \notag\\
				Y_2 +Y_8+Y_{14}& \xrightarrow{\theta_3} 	Y_4+Y_{10} +Y_{16} \notag\\
				Y_1+Y_7+Y_{13} & \xrightarrow{\theta_4}  	Y_4 +Y_{10}+Y_{16}\notag\\
				Y_4+Y_{10}+Y_{16}& \xrightarrow{\theta_5}  	Y_6+Y_{12}+Y_{18} \notag\\
				Y_5+Y_{11}+Y_{17} & \xrightarrow{\theta_6}  	Y_5+Y_{11} +Y_{17}\notag
			\end{cases}	
		\end{align}
	\end{minipage}	
	\caption{Three dynamic systems with $r=6$ reactions, and an increasing number of particles ($p=6, 12, 18$).}
	\label{var p}
\end{table}

 \begin{table}[h!]
 	\tiny
 	\centering
 	\begin{minipage}{.2\textwidth}
 		\centering
 		$r=6$
 		\begin{align}
 			\mathcal{R}_6: 
 			\begin{cases}
 				Y_2 & \xrightarrow  {\theta_1} Y_1+Y_3 \notag\\
 				Y_3 & \xrightarrow{\theta_2} 	Y_2+Y_4  \notag\\
 				Y_4 & \xrightarrow{\theta_3} 	Y_3+Y_5  \notag\\
 				Y_5 & \xrightarrow{\theta_4}  	Y_4+Y_6 \notag\\
 				Y_5& \xrightarrow{\theta_5}  	Y_6 \notag\\
 				Y_6 & \xrightarrow{\theta_6}  	Y_1 \notag
 			\end{cases}
 		\end{align}
 	\end{minipage}
 	\hspace{1cm}
 	\begin{minipage}{.2\textwidth}
 		\centering
 		$r=12$
 		\begin{align}
 			\mathcal{R}_{12}:	\mathcal{R}_6 \cup 
 			\begin{cases}
 				Y_8 & \xrightarrow{\theta_1} Y_7+Y_9 \notag\\
 				Y_9 & \xrightarrow{\theta_2} Y_8+Y_{10} \notag\\
 				Y_{10} & \xrightarrow{\theta_3} Y_9+Y_{11} \notag\\
 				Y_{11} & \xrightarrow{\theta_4} Y_{10}+Y_{12} \notag\\
 				Y_{12} & \xrightarrow{\theta_5} Y_11 \notag\\
 				Y_7 & \xrightarrow{\theta_6} Y_12 \notag
 			\end{cases}
 		\end{align}
 	\end{minipage}	
 	\hspace{1cm}
 	\begin{minipage}{.2\textwidth}
 		\centering
 		$r=18$
 		\begin{align}
 			\mathcal{R}_{18}:	\mathcal{R}_{12} \cup 
 			\begin{cases}
 				Y_{14} & \xrightarrow{\theta_1} Y_{13}+Y_{15} \notag\\
 				Y_{15} & \xrightarrow{\theta_2} Y_{14}+Y_{16} \notag\\
 				Y_{16} & \xrightarrow{\theta_3} Y_{15}+Y_{17} \notag\\
 				Y_{17} & \xrightarrow{\theta_4} Y_{16}+Y_{18} \notag\\
 				Y_{18} & \xrightarrow{\theta_5} Y_{17} \notag\\
 				Y_{13} & \xrightarrow{\theta_6} Y_{18} \notag
 			\end{cases}
 		\end{align}
 	\end{minipage}	
 	\caption{Three dynamic systems with $p=6$ particles, and an increasing number of reactions ($r=6, 12, 18$).}
 	\label{var r}
 \end{table} 

\bibliographystyle{chicago}
\bibliography{biblio}

\end{document}